\documentclass[aps,prd,twocolumn,preprintnumbers,superscriptaddress,nofootinbib,,10pt,showkeys]{revtex4-2}
\usepackage{hyperref}
\usepackage{amsmath}

\usepackage{newtxtext,newtxmath}
\usepackage[T1]{fontenc}

\DeclareRobustCommand{\VAN}[3]{#2}
\let\VANthebibliography\thebibliography
\def\thebibliography{\DeclareRobustCommand{\VAN}[3]{##3}\VANthebibliography}

\usepackage{graphicx}	
\usepackage{dcolumn}
\usepackage{bm}
\usepackage{array}

\usepackage[dvipsnames]{xcolor}

\newcommand{\ie}{i.e.,~}
\newcommand{\eg}{e.g.,~}

\newcommand*{\rom}[1]{\expandafter\@slowromancap\romannumeral #1@}

\begin{document}

\title{General gravitational properties of neutron stars:\\ curvature
  invariants, binding energy, and trace anomaly}

\author{Iv\'an Garibay}\thanks{Contact author: garibay@itp.uni-frankfurt.de}
\affiliation{Institut f\"ur Theoretische Physik, Goethe Universit\"at,
 Max-von-Laue-Str. 1, 60438 Frankfurt am Main, Germany}
\affiliation{Instituto de Astronom\'ia, Universidad Nacional Aut\'onoma de M\'exico,
 Ciudad de M\'exico, CDMX 04510, Mexico}

\author{Christian Ecker}
\affiliation{Institut f\"ur Theoretische Physik, Goethe Universit\"at,
 Max-von-Laue-Str. 1, 60438 Frankfurt am Main, Germany}

\author{Luciano Rezzolla}
\affiliation{Institut f\"ur Theoretische Physik, Goethe Universit\"at,
 Max-von-Laue-Str. 1, 60438 Frankfurt am Main, Germany}
\affiliation{School of Mathematics, Trinity College, Dublin 2, Ireland}
\affiliation{Frankfurt Institute for Advanced Studies, Ruth-Moufang-Str. 1,
 60438 Frankfurt am Main, Germany}

\date{\today}

\begin{abstract}
\noindent 
We investigate the behavior of curvature invariants for a large ensemble
of neutron stars built with equations of state (EOSs) that satisfy
constraints from nuclear theory and perturbative QCD, as well as
measurements of neutron-star masses, radii, and gravitational waves from
binary neutron-star mergers.
Surprisingly, our analysis reveals that stars with negative Ricci scalar
$\mathcal{R}$ are rather common and about $\sim 50\%$ of our EOSs produce
one or more stars with Ricci curvature that is negative somewhere inside
the star. The negative curvature is found mostly but not exclusively at
the highest densities and pressures, and predominantly for stiff EOSs and
for the most compact and most massive stars. Furthermore, we improve the
quasi-universal relation between the stellar gravitational mass $M$ and
the baryonic mass $M_\mathrm{b}$, which allows us to express analytically
one in terms of the other with a maximum variance of only $\sim 3\%$.
Finally, using the relation between the Ricci scalar and the trace
anomaly $\Delta$, we determine the conditions under which $\Delta$
vanishes or becomes negative in neutron stars.
\end{abstract}

\keywords{neutron stars -- equation of state -- trace anomaly --
  curvature scalars}

\maketitle

\section{Introduction}
\label{sec:intro}

The extreme conditions inside neutron stars (NSs) present a major
challenge for the theoretical modelling of nuclear matter at densities
several times higher than the saturation density of atomic nuclei, $n_s
:= 0.16\,\rm fm^{-3}$. While effective field theory remains one of the
most powerful tools for predicting the behaviour of dense matter, its
uncertainties grow significantly at the high densities characteristic of
NS cores. On the other hand, first-principles perturbative Quantum
Chromodynamics (QCD) calculations are reliable only at densities far
exceeding those found in NSs, but they still offer valuable consistency
checks for modelling matter at lower densities~\citep{Fraga2014,
  Most2018, Annala2019, Ferreira2021a, Komoltsev:2021jzg, Annala:2022,
  Somasundaram:2022, Gorda:2022jvk, Fujimoto2023, Albino2024,
  Fukushima2025}. As a result, our theoretical understanding of
fundamental properties -- such as the equation of state (EOS) of dense
nuclear and quark matter -- remains limited. To counter this and make
progress on our understanding of NSs, agnostic or phenomenological
approaches are employed to construct the EOS at NS densities. At the same
time, recent and forthcoming observations of NSs and their mergers offer
a unique opportunity to probe strongly coupled dense matter under
conditions that are otherwise inaccessible in terrestrial experiments.

It is therefore essential to integrate existing theoretical insights with
observational data to constrain the EOS and related quantities that
govern the macroscopic properties of NSs. Thus far, the primary focus of
this effort has been on the EOS itself, with significant progress
achieved through the use of generic parametrizations combined with
statistical methods to derive robust constraints on the distribution of
masses and radii from theoretical and astrophysical
inputs~\citep{Coughlin2017}. This has been followed by more targeted
studies aimed at constraining quantities closely tied to the EOS, such as
the speed of sound~\citep{Moustakidis2017, Tews2018a, Margaritis2020,
  Kanakis-Pegios:2020kzp, Hippert:2021, Altiparmak:2022} and the trace
(conformal) anomaly~\citep{Fujimoto:2022ohj_pub, Marczenko:2022jhl,
  Ecker:2022b, Annala2023, Marczenko:2025hsh, Ecker2025}.

However, in addition to these physically measurable quantities, the use
of parameterized EOSs also allows one to explore more geometrical
properties of NSs that, while not measurable, provide important insights
in the gravitational properties of these objects and, ultimately, about
gravity under extreme conditions (see, \eg
  \cite{Ecker:2024}). Given the general information about curvature that
is accessible via the Riemann tensor, the curvature scalars that can be
constructed from it and its variants, such as the Ricci scalar
$\mathcal{R}$ and the Kretschmann scalar $\mathcal{K}_1$, provide
quantitative information about the internal geometry of NSs.

In this context, interesting studies on curvature scalars and their role
in alternative theories of gravity, dark-matter components, and
anisotropic EOSs have already been presented in the literature~\citep[see,
  \eg][]{Eksi2014, He2015, Das2021, Biswal2025, Danarianto2025,
  Ghosh2025, Khunt2025}. All of these studies, however, have concentrated
on specific EOSs and have not really explored the behaviour of
curvature-related quantities in NSs within a comprehensive,
  statistical, and agnostic description of the EOSs. As a result, the
existence of quasi-universal behaviour of geometry and gravitationally
related quantities has not been explored yet and represents the main goal
of this work. More specifically, we here present a systematic
investigation of the behaviour within the star of various curvature
invariants and highlight the conditions under which they can become
negative. Furthermore, we statistically improve previous studies on the
quasi-universal relation between the baryonic and gravitational mass,
compactness and binding energy of NSs. Finally, we relate the properties
of the Ricci scalar with those of the trace anomaly and illustrate the
non-trivial correlation that exists between the two as a function of the
stellar mass.

The plan of the paper is as follows. In Sec.~\ref{sec:methods} we briefly
describe the numerical implementation of our set of constrained
parametric EOSs. Additionally, we include the mathematical formalism for
the curvature scalars and their relation with the matter content inside
NSs. In Sec.~\ref{sec:results} we present and discuss our
results, while Sec.~\ref{sec:conclusion} is dedicated to our conclusions
and prospects for future research. Geometric units in which $c=G=1$ are
used throughout this work and we adopt the standard notation for the
summation on repeated tensor indices, which we take to run from $0$ to
$3$ and indicate with Greek letters.

\section{Methods}
\label{sec:methods}  

\subsection{Stellar Equilibrium Models}

To construct static and spherically symmetric stellar models in
hydrostatic equilibrium we assume the matter to be described by a perfect
fluid with energy-momentum tensor~\citep{Rezzolla_book:2013}
\begin{equation}
T^{\mu \nu} = \left( e + p \right) u^\mu u^\nu  + pg^{\mu \nu} \, ,
\label{eq:Tmunu}
\end{equation}
where $u^{\mu}$ are the components of the fluid four-velocity, 
$g^{\mu \nu}$ is the metric tensor, and $p$
and $e$ are the pressure and energy density, respectively.

Einstein's equations, together with the conservation of the
energy-momentum tensor, reduce to the relativistic hydrostatic
equilibrium equation
\begin{equation}
  \label{eq:TOV_P}
 \frac{dp}{dr} = - \frac{m e}{r^2} \left( 1 + \frac{p}{e} \right)
 \left( 1 + \frac{ 4\pi r^3 p}{m} \right) \left( 1-\frac{2m}{r} \right)
^{-1} \, ,
\end{equation}
where $r$ is the radial coordinate, while the integrated mass $m = m(r)$
enclosed within $r$ follows the differential equation
\begin{equation}
  \label{eq:TOV_m}
 \frac{dm}{dr} = 4\pi r^2 e \, .
\end{equation}
Equations~\eqref{eq:TOV_P}--\eqref{eq:TOV_m} are also known as the
Tolman-Oppenheimer-Volkov (TOV) equations.

Given an EOS, $p = p(e)$, a unique star with stellar radius $R$ and
stellar mass $M$ is defined imposing the set of boundary conditions at
the centre of the star, $e (0) = e_c$ and $m(0) = 0$; and at the surface
of the star $r=R$, where $p(R) = 0$ and $M := m(R)$. The maximum stable
mass achieved by an EOS will be denoted hereafter by $M_{_{\mathrm{TOV}}}$.

\subsection{Equations of State}

Our construction of the EOSs follows closely the framework
of~\citet{Altiparmak:2022}, and we thus generate a large ensemble by
combining established models in different density regimes: the
Baym-Pethick-Sutherland (BPS) model~\citep{Baym71b} for the crust,
polytropic segments sampled between soft and stiff nuclear-theory
bounds~\citep{Hebeler:2013nza} up to $1.1\,n_s$, and perturbative QCD
results~\citep{Fraga2014} at baryon chemical potentials above
$2.6\,\mathrm{GeV}$. In the intermediate density regime relevant for NSs,
we adopt the piecewise-linear sound-speed parametrization introduced
by~\citet{Annala2019}, where $c^2_s := \partial p / \partial e $ is the
speed of sound. This approach allows one to obtain a smooth interpolation
between nuclear and quark matter. For our Bayesian inference we use a
uniform prior distribution of our parameters -- the piecewise speeds of
sound with $N=7$ segments -- within the allowed interval,
$c_{\mathrm{s},i}^2 \in [0,1]$.

To constrain this set, we retain only EOSs compatible with astrophysical
observations. These include: the existence of $\sim2\,M_\odot$
NSs~\citep{Antoniadis2013,Cromartie2019,Fonseca2021}, NICER radius
measurements of pulsars J0740+6620 and J0030+0451~\citep{Miller2021,
  Riley2021, Riley2019, MCMiller2019b}, and the tidal deformability
constraint from the gravitational waves event
GW170817~\citep{Abbott2018a}. Specifically, we impose the following
cut-off filters through our likelihood function: $M_{\rm TOV} >
2.08\,M_\odot$; radii $R>10.75\,\mathrm{km}$ for $2.0\,M_\odot$ NSs;
$R>10.8\,\mathrm{km}$ for $1.1\,M_\odot$ NSs; and a tidal deformability
$\tilde\Lambda < 720$ for a binary of NSs with chirp mass
$\mathcal{M}_{\rm chirp} = 1.186\,M_\odot$ and mass ratio $q>0.73$.
Varying the values of our parameters with a Goodman and Weare's Affine
invariant Markov chain Monte Carlo ensemble sampler \citep{Goodman2010}
\footnote{From the Python package \texttt{emcee}.}, we obtain a posterior
of approximately $3 \times 10^4$ viable EOSs.

\subsection{Curvature Invariants}

We recall that in general relativity, the curvature of spacetime is
encoded in the Riemann tensor. A natural approach to investigate the
connection between the properties of the EOS and the spacetime curvature
in NSs consists of considering coordinate (gauge)–independent measures of
curvature, namely scalar curvature invariants, which are constructed by
contracting tensor products of the Riemann tensor with itself and other
geometric objects. While there are 14 independent second-order curvature
scalar invariants~\citep{Geheniau1956}, in this work we focus on the
Ricci scalar, together with the three independent quadratic curvature
invariants formed from contractions of the Riemann tensor and the
Levi-Civita tensor. Our discussion here closely
follows~\cite{Ecker:2024}, where the same scalars are analyzed in the
context of binary NS mergers for a small set of EOSs with phase
transitions~\citep{Demircik:2021zll}.

More specifically, using the Ricci tensor, $\mathcal{R}_{\mu \nu}$, it is
possible to compute its full contraction
\begin{align}
\mathcal{J}^2 := \mathcal{R}_{\mu \nu} \mathcal{R}^{\mu \nu} \,.
\end{align}
as well as its contraction with the metric tensor $g^{\mu \nu}$, which
yields the well-known Ricci scalar
\begin{equation}
\mathcal{R} := \mathcal{R}_{\mu \nu} g^{\mu \nu} = \mathcal{R}^{\mu}_{~\mu}\,,
\end{equation}
which plays a fundamental role in Einstein equations.

Other quadratic invariants -- known as the Kretschmann scalar
$\mathcal{K}_1$, the Chern-Pontryagin scalar $\mathcal{K}_2$, and the
Euler scalar $\mathcal{K}_3$ -- constitute the principal invariants of
the Riemann tensor on four-dimensional Lorentzian
manifolds~\citep{Cherubini:2002gen}, and are defined as follows:
\begin{align}
 \mathcal{K}_1&:=\mathcal{R}_{\mu \nu \rho \sigma}\,\mathcal{R}^{\mu \nu
   \rho \sigma}\,,\\
 \mathcal{K}_2&:={{}^\star \mathcal{R}}_{\mu \nu \rho
   \sigma}\,\mathcal{R}^{\mu \nu \rho \sigma}\,,\\
 \mathcal{K}_3&:={{}^\star \mathcal{R}^\star}_{\mu \nu \rho
   \sigma}\,\mathcal{R}^{\mu \nu \rho \sigma}\,,
\end{align}
where the left and right duals of a generic four-index tensor
$\mathcal{A}_{\mu \nu \rho \sigma}$ are defined via contractions with the
Levi-Civita tensor $\epsilon_{\mu \nu \alpha \beta}$:
\begin{align}
 {{}^\star \mathcal{A}}_{\mu \nu \rho \sigma}&:=\epsilon_{\mu \nu \alpha
   \beta}{\mathcal{A}^{\alpha \beta}}_{\rho \sigma}\,,\\
 \mathcal{A}^\star_{\mu \nu \rho \sigma} &:={\mathcal{A}_{\mu
     \nu}}^{\alpha \beta}\epsilon_{\alpha \beta \rho \sigma}\,.
\end{align}

Note that $\mathcal{K}_2$ can equivalently be defined using the right
dual instead of the left dual; the difference is irrelevant for our
purposes.

Using the Weyl tensor, that is, the trace-free part of the Riemann tensor
\begin{align}
  \mathscr{W}_{\mu \nu \rho \sigma} := \mathcal{R}_{\mu \nu \rho \sigma}
   -\left(g_{\mu[\rho}\mathcal{R}_{\sigma]\nu} -
   g_{\nu[\rho}\mathcal{R}_{\sigma]\mu}\right) +
   \frac{1}{3}\mathcal{R}~g_{\mu[\rho}g_{\sigma]\nu}\,,
\end{align}
where the square brackets denote antisymmetric tensor indices, it is
possible to compute two additional invariants defined as
\begin{align}
  &\mathcal{I}_1 := \mathscr{W}_{\mu \nu \rho \sigma}\mathscr{W}^{\mu \nu
    \rho \sigma}\,,\\
  &\mathcal{I}_2:={{}^\star \mathscr{W}}_{\mu \nu \rho
    \sigma}\,\mathscr{W}^{\mu \nu \rho \sigma}\,.
\end{align}
These are the only linearly independent principal invariants of the 
four-dimensional Weyl tensor, since
\begin{equation}
{{}^\star\mathscr{W}^\star}_{\mu \nu \rho \sigma}\,\mathscr{W}^{\mu \nu
  \rho \sigma} = -\mathcal{I}_1\,.
\end{equation}

The invariants introduced so far are not linearly independent and are
indeed related via the following equations~\citep{Cherubini:2002gen}
\begin{align}
\label{eq:K_1}
 \mathcal{K}_1&=\mathcal{I}_1+2\mathcal{J}^2-\frac{1}{3}\mathcal{R}^2\,,\\
\label{eq:K_2}
 \mathcal{K}_2&=\mathcal{I}_2\,,\\
\label{eq:K_3}
 \mathcal{K}_3&=-\mathcal{I}_1+2\mathcal{J}^2-\frac{2}{3}\mathcal{R}^2\,.
\end{align}
Furthermore, since we are considering static, spherically
symmetric, and asymptotically flat spacetimes, $\mathcal{K}_2$ -- and
hence $\mathcal{I}_2$ -- are identically zero~\citep{Plebanski1968}.

\subsection{Energy-Momentum Tensor and Trace-Anomaly}

Given the energy-momentum tensor of a perfect fluid~\eqref{eq:Tmunu} and
a timelike 4-velocity $u^\mu$ with unit norm $u^\mu u_\mu = -1$, the
trace of the energy-momentum tensor is simply given by
\begin{equation}
  \label{eq:trace_Tmunu}
  T := T^{\mu\nu} g_{\mu\nu} = 3p-e \, .
\end{equation}
We can now use Eq.~\eqref{eq:trace_Tmunu} to introduce a basic diagnostic
of the EOS behavior, namely, the so-called trace (or ``conformal'')
anomaly\footnote{The ``conformal'' nomenclature comes from the fact that
for a conformal fluid $p=e/3$, so that $\Delta \neq 0$ measures the
anomaly with respect to a conformal fluid.}, which is defined as
\begin{equation}
  \Delta := \frac{1}{3} - \frac{p}{e}  = -\frac{T}{3e}\,.
\end{equation}
Note that since $e\geq0$, then $T\leq0$ ($T>0$) when $\Delta \geq 0$
($\Delta <0$).

This dimensionless quantity measures deviations from conformal symmetry
and is bounded by causality and thermodynamic stability within $-2/3 \leq
\Delta \leq 1/3$. It is worth mentioning that the sector $\Delta < 0$
does not imply violation of causality, since theoretical relativistic
EOSs can be constructed with $\Delta < 0$, as noted
by~\citet{Zeldovich1961}. Obviously, causality also requires a subluminal
speed of sound of the EOS, $\ie 0 \leq c_s \leq 1$. In the context of
model-agnostic parametrizations of NS matter, a conformal anomaly
approaching zero has been used as an indicator of a deconfinement
transition to quark matter~\cite{Annala2023, Blomqvist2025}, a trend that
has been confirmed by microphysical modelling of
deconfinement~\cite{Ecker2025}.

For a static and spherically symmetric perfect fluid, it is possible to
write explicit relations between the trace anomaly and some of
the curvature scalars, \ie $\mathcal{R}$, $\mathcal{J}^2$, and
$\mathcal{I}_1$. More specifically, we introduce the following
normalised quantities
\begin{align}
\label{eq:mean_e}
 \bar{e} &:= \frac{3m}{{4} \pi r^3} \,, \\
\label{eq:Psi}
 \Psi &:= 1- \frac{\bar{e}}{e} \,,
\end{align}
where $\bar{e}$ is the mean density enclosed within a sphere of
coordinate radius $r$ and $\Psi$ measures the deviation of the energy
density $e$ from $\bar{e}$. In this way, the following expressions
can be obtained
\begin{align}
\label{eq:Ricci}
 \mathcal{R} &= 24\, \pi e \Delta \,, \\
\label{eq:J}
 \mathcal{J}^2 &= 64\, \pi^2 e^2 \left[ 1 + 3\left( \frac{1}
     {3} - \Delta \right)^2 \right] \,, \\
\label{eq:W}
 \mathcal{I}_1 &= \frac{256}{3} \pi^2 e^2\, \Psi ^2 = \frac{256}{3} \pi^2
 \left( \frac{3m}{{4} \pi r^3} - e \right)^2\,.
\end{align}

\begin{figure*}
\centering
\includegraphics[width=0.99\textwidth]{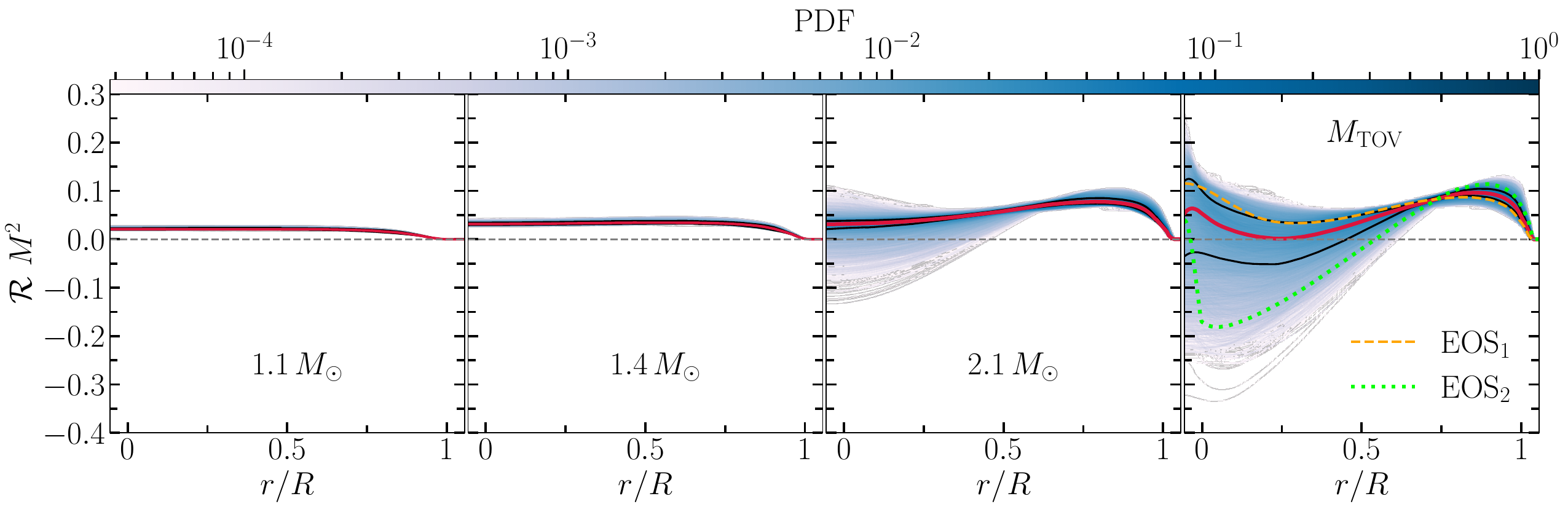}
\caption{Normalised PDFs of the radial profiles of the normalised Ricci
  scalar $\mathcal{R} \, M^2$ for fixed NS masses. The red lines mark the
  median values of the distribution, while black lines correspond to
  $1$-$\sigma$ confidence limits. From left to right: the profiles refer
  to NSs with masses $1.1 \, M_\odot$, $1.4 \, M_\odot$, $2.1 \, M_\odot$
  and $M_{_{\mathrm{TOV}}}$, respectively. Shown in the rightmost
    panel are two reference EOSs: ${\rm EOS}_{1}$ (orange dashed) and
    ${\rm EOS}_{2}$ (green dotted line).}
\label{fig:fig1}
\end{figure*}

Since the solution of the TOV
equations~\eqref{eq:TOV_P}--\eqref{eq:TOV_m} amounts to computing the
(radial) functional behaviour of $e=e(r)$, $p=p(r)$, and $m=m(r)$, we can
use the relations (\ref{eq:K_1})-(\ref{eq:K_3}), together with
(\ref{eq:Ricci})-(\ref{eq:W}), to derive the radial profile of the
curvature scalars $\mathcal{K}_1 = \mathcal{K}_1 (r)$ and $\mathcal{K}_3
= \mathcal{K}_3 (r)$ inside the NSs. Finally, another useful scalar
quantity whose behaviour inside the star can be of interest is simply
given by the local stellar compactness
\begin{equation}
  \label{eq:compactness}
  \eta(r) := \frac{m(r)}{r} \, ,
\end{equation}
which increases from zero to its maximum value $\eta_{\rm max}$ close to
the crust~\citep{Eksi2014}\footnote{We note that our definition of
compactness differs by a factor of two from that given
by~\cite{Eksi2014}.}. It should be noted that $\eta_{\rm max}$ is in
general larger than the asymptotic value of $\eta(r)$, \ie $\mathscr{C}
:= \lim_{r\to \infty} \eta(r) = M/R$, which is also referred to as the
(asymptotic) stellar compactness. It has been recently argued that
$\mathscr{C} \leq 1/3$ for realistic NSs~\citep{Rezzolla2025} and indeed
all of our stellar models are consistent with this upper bound.

\section{Results}
\label{sec:results}

\subsection{Ricci Scalar}

Figure~\ref{fig:fig1} presents with a colormap the radial profiles of
probability distribution functions (PDFs) for the Ricci scalar relative
to all of the stellar models considered and for every EOS. Since the
distributions vary considerably with the mass, we report in the first
panels from left the radial profiles for a fixed mass of $1.1, \, 1.4,$
and $2.1 \, M_\odot$; the last and fourth panel, on the other hand,
collects stellar models at the maximum mass $M_{_{\mathrm{TOV}}}$, which
have larger variation since different EOSs will in general feature
significantly different maximum masses. Note that the plots refer to
dimensionless quantities since the Ricci scalar is scaled by the square
of the NS mass, $\mathcal{R} M^2$, and the radial position is rescaled by
the corresponding stellar radius, $r/R$. The red solid line represents
the median-value of the distribution, while the black solid lines
correspond to the $1$-$\sigma$ deviation from it. In the rightmost
  panel, corresponding to the $M_\mathrm{TOV}$ star, reported with an
  orange dashed and a green dotted line are the radial profiles of the
  normalised Ricci scalar for the representative EOSs, $\mathrm{EOS}_1$
  and $\mathrm{EOS}_2$, respectively.

\begin{figure*}
  \centering
  \includegraphics[width=0.45\textwidth]{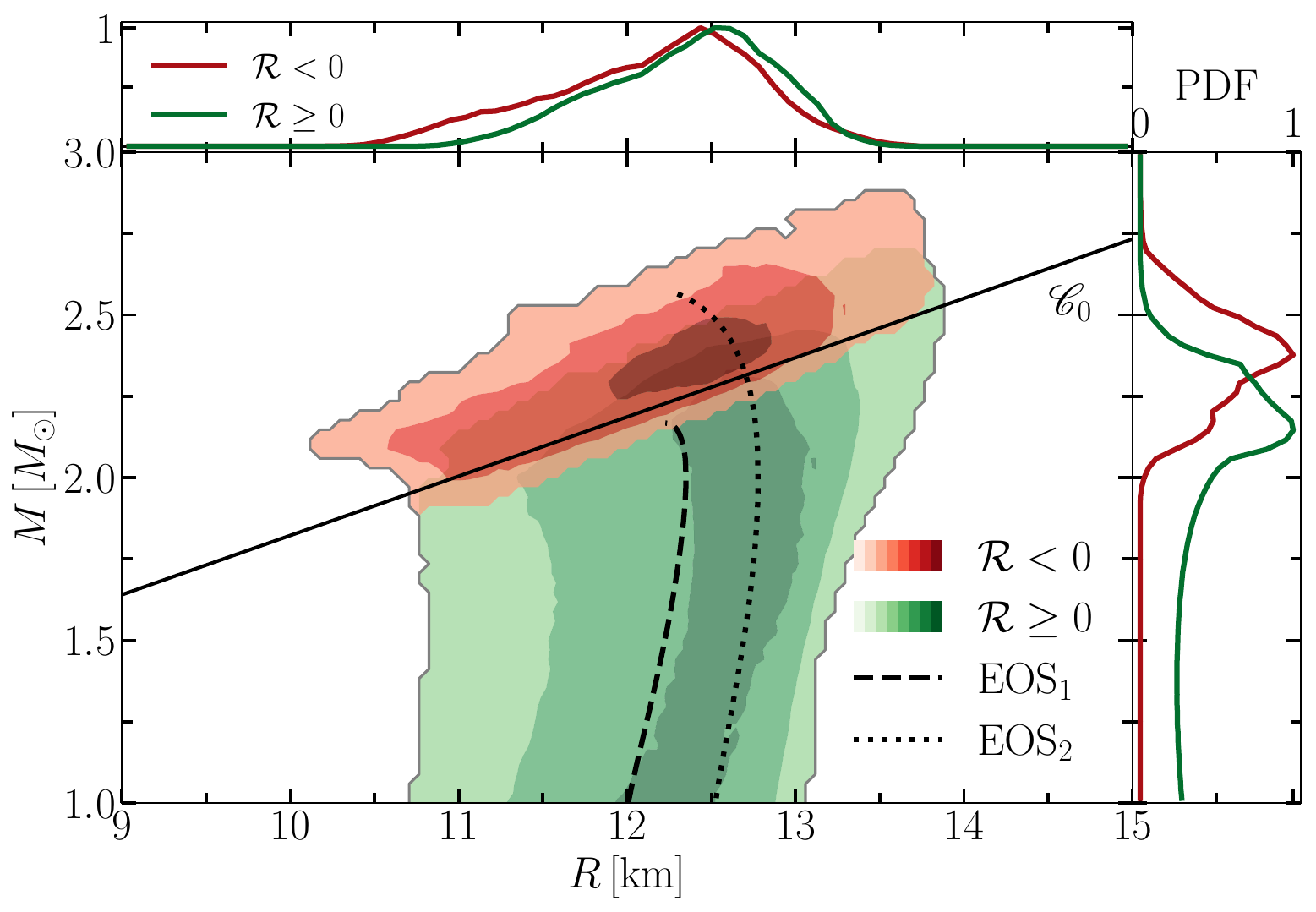}
  \hspace{1.0cm}
  \includegraphics[width=0.45\textwidth]{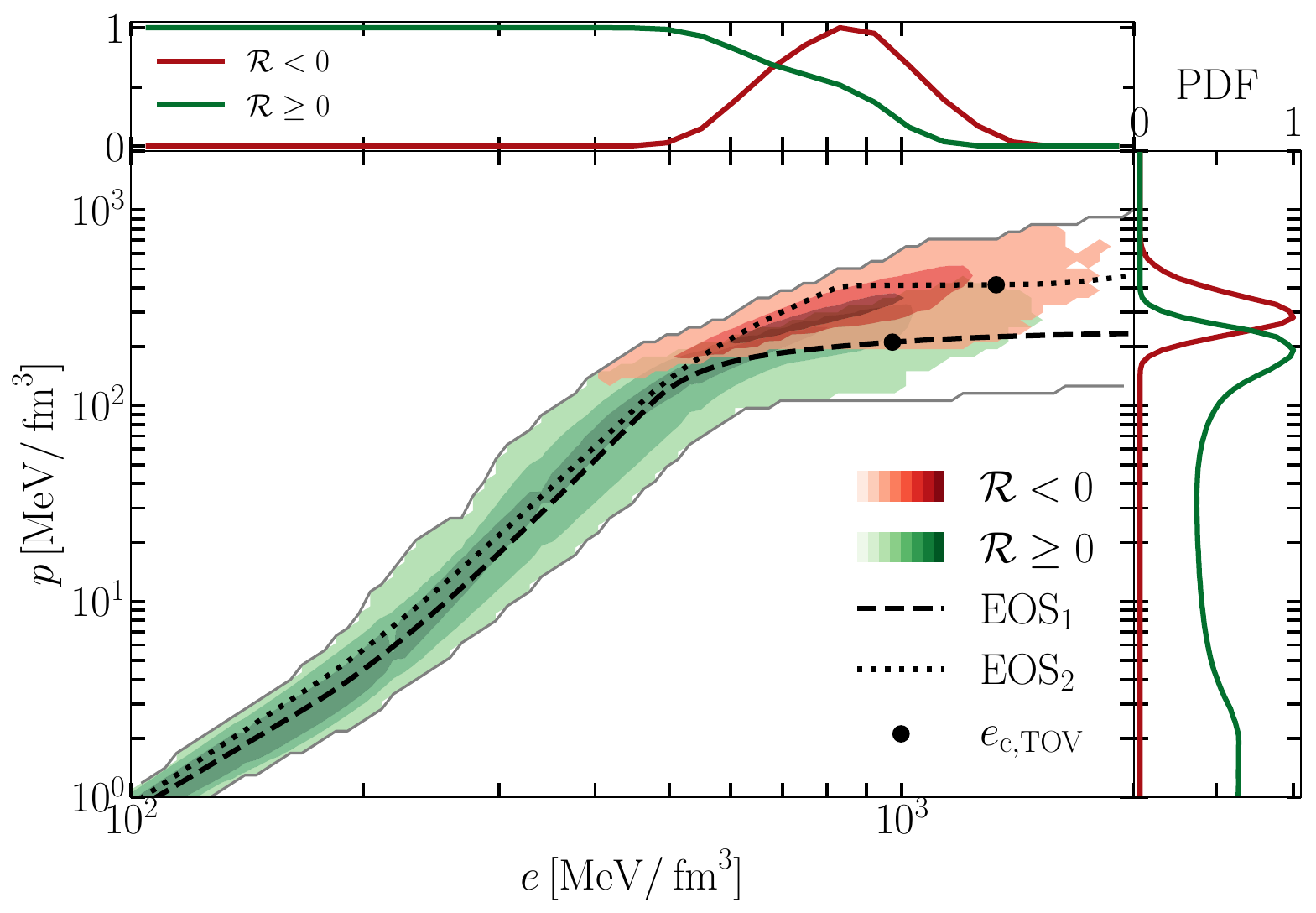}
  \caption{\textit{Left panel:} Normalised PDFs of the $M$-$R$ relations
    for the entire EOS ensemble. The green/red colormap represents the
    distribution of NSs with non-negative/negative Ricci scalar,
    everywhere/somewhere inside the star. $1$-$\sigma$, $2$-$\sigma$ and
    $100\%$-confidence intervals are shown from darker to lighter
    colours. The grey solid line contours the $100\%$-confidence interval
    for every stable NS. Shown with a black solid is the
    ``zero-curvature'' compactness relation $\mathscr{C}_0$, while black
    dashed and dotted lines mark two reference EOSs (${\rm
      EOS}_{1,2})$. The top and right parts of the panel report the
    $100\%$-confidence intervals of both distributions in the relevant
    direction. \textit{Right panel}: the same as in the left but for the
    $p$-$e$ relations. Marked with black filled circles are the largest
    energy densities for stable NSs with ${\rm EOS}_{1,2}$.}
  \label{fig:fig2}
  \end{figure*}

When concentrating for simplicity on the median values, the first two
panels from the left in Fig.~\ref{fig:fig1} clearly show that the Ricci
scalar is always positive and essentially constant across the star,
retaining the value at the centre up to the outer layers of the star,
where $\mathcal{R} \to 0$, as the stellar surface is approached. At the
same time, when considering the mass-cut at $2.1\,M_{\odot}$ (third
panel), it is possible to note that the Ricci scalar is no longer
monotonic and that it actually increases to a local maximum around $r/R
\simeq 0.7$ to then decrease towards the surface. Also, while the average
value is always positive, a small portion of the EOSs does show stellar
models with negative Ricci scalar in the stellar core. This behaviour is
magnified when considering a mass cut at $M=M_{\mathrm{TOV}}$, as shown
in the fourth (rightmost) panel of Fig.~\ref{fig:fig1}. In this case, in
fact, it is possible to note that $\mathcal{R}$ is not only
non-monotonic, but that it can have negative values in the inner regions
of the stars. Indeed, while the median value of $\mathcal{R} > 0$ for
$r/R \simeq 0$, it vanishes for $r/R \simeq 0.25$. The Ricci scalar then
grows to a local maximum near the stellar surface and naturally vanishes
as the stellar surface is approached. It is of interest to remark
  that the radial profile corresponding to $\mathrm{EOS}_2$ presents a
  jump close to the core, as it transits from negative values to positive
  ones at the center. This is due to the appearance in $\mathrm{EOS}_2$
  of a first-order phase transition, as it can be appreciated
  on the right panel of Fig.~\ref{fig:fig2}, where the pressure exhibits
  a plateau at high densities before reaching the central energy density
  of the TOV NS, leading to a sound speed close to zero. By contrast,
  $\mathrm{EOS}_1$ presents a smoother radial profile of the Ricci scalar
  given that it does not exhibit a first-order phase transition at these
  densities.

The left panel of Fig.~\ref{fig:fig2} displays with two colormaps the
PDFs of the Ricci scalar as a function of the mass and radii of the
various stellar models. More specifically, while the green colormap is
used for those stellar models having non-negative Ricci scalar, \ie
$\mathcal{R} \geq 0$, everywhere inside the star, the red colormap
corresponds to those NSs with negative Ricci scalar, \ie $\mathcal{R} <
0$, somewhere inside the star. In both cases, $1$-$\sigma$, $2$-$\sigma$
confidence levels are shown with darker and lighter shades of the
colours, while the solid grey line marks the $100\%$-confidence interval.
The distributions clearly show that more compact stars are more likely to
have negative values of the Ricci scalar and this can happen either at
small radii ($R\simeq 11\,{\rm km}$) and with comparatively small masses
($M \gtrsim 2.0\, M_{\odot}$) or at larger radii ($R\simeq 13\,{\rm km}$)
and with comparatively large masses ($M \gtrsim 2.3\, M_{\odot}$). Also
worth remarking is that the most compact stars -- and hence massive stars
for those EOSs~\citep[see][for a conjecture relating the maximum
  compactness to the maximum mass]{Rezzolla2025} -- all have a negative
Ricci scalar somewhere in their interior.

Note also that the two regions with $\mathcal{R} \lessgtr 0$ are not
distinct but they overlap. This can be appreciated from the projections
of the $100\%$-confidence intervals of the distribution as a function of
the stellar radius (red and green solid lines on the top part of the
panel) and of the stellar mass (red and green solid lines on the right
part of the panel). The local maxima at $M \sim 2.3\,M_\odot$ and $R
\sim 12.5\,{\rm km}$ simply reflect a higher concentration of stellar
models in those regions~\citep{Altiparmak:2022}.

From our collection of EOSs, about $51\%$ of them leads to NSs with
$\mathcal{R}>0$ everywhere in their interior. While the rest of EOSs
produce at least one NS with $\mathcal{R}<0$ somewhere in their
interior. We note that, for every EOS, it is possible to determine the
``zero-curvature'' compactness $\mathscr{C}_0$, that is, the ratio $M/R$
at which the stellar model has zero Ricci curvature at the centre, \ie
$\mathcal{R}(r=0) = 0$\footnote{After fixing the EOS, it is possible that
two NSs have $\mathcal{R}(r=0) = 0$ (see also the discussion on
Fig.~\ref{fig:fig6} below); when this happens, and in order to compare
our results with those of~\cite{Podkowka2018}, we consider the least
massive of the two stars.}. Interestingly, it was shown in~\cite{Podkowka2018}
using 136 piecewise polytropic EOSs that this
compactness is quasi-universal, \ie does not depend significantly on the
EOS and is given by $\mathscr{C}_0 =
0.262^{+0.011}_{-0.017}$~\citep{Podkowka2018}.

We have found that this result is very robust and indeed our estimate
obtained on the basis of the $10^4$ EOSs of our sample differs only by
$3\%$ and is given by $\mathscr{C}_0 = 0.26911 \pm 0.00004$. We report
this ``zero-curvature'' compactness as a black solid line in the left
panel of Fig.~\ref{fig:fig2}. The right panel of Fig.~\ref{fig:fig2}
provides information that is similar and complementary to that reported
in the left panel. More specifically, using the same notation, the panel
reports the PDFs of the Ricci scalar as a function of the pressure and
energy density of the various EOSs. It is therefore not surprising that
the stiffer EOSs tend to lead to NSs with $\mathcal{R} < 0$. Less
obvious, however, is the fact that at the highest densities and
pressures, all stellar models have negative Ricci scalar in their
interior. Overall, the results in Fig.~\ref{fig:fig2} clearly show that
the most compact and most massive stars must have $\mathcal{R} < 0$. As
we discuss further below, this has direct implications on the the trace
anomaly.

\subsection{Gravitational and Baryonic Mass}

Measured gravitational wave (GW) signals and light curves from black
hole-NS or NS-NS mergers can shed light on the binary
properties~\citep[see, \eg][for some reviews]{Baiotti2016,
  Paschalidis2016, Radice2020b}. The gravitational mass of a NS, $M$, is
the value obtained by integrating Eq.~\eqref{eq:TOV_m} from the center of
the star to its surface:
\begin{equation}
 M := \int_{0}^{R} \frac{dm}{dr} dr \,,
\end{equation}
while the baryonic mass of a NS, $M_\mathrm{b}$, is the value obtained by
integrating from the center to the surface the following quantity:
\begin{equation}
  M_\mathrm{b} := 4\pi m_\mathrm{u} \int_{0}^{R} \frac{n_\mathrm{b} r^2
    dr}{\sqrt{1-{2m(r)}/{r}}} \,,
\end{equation}
where $m_\mathrm{u}\approx 8.3486\times10^{-58}~M_\odot$ is the atomic
mass unit, and $n_\mathrm{b}$ is the baryon number density.

\begin{figure*}
  \centering
  \includegraphics[width=0.4\textwidth]{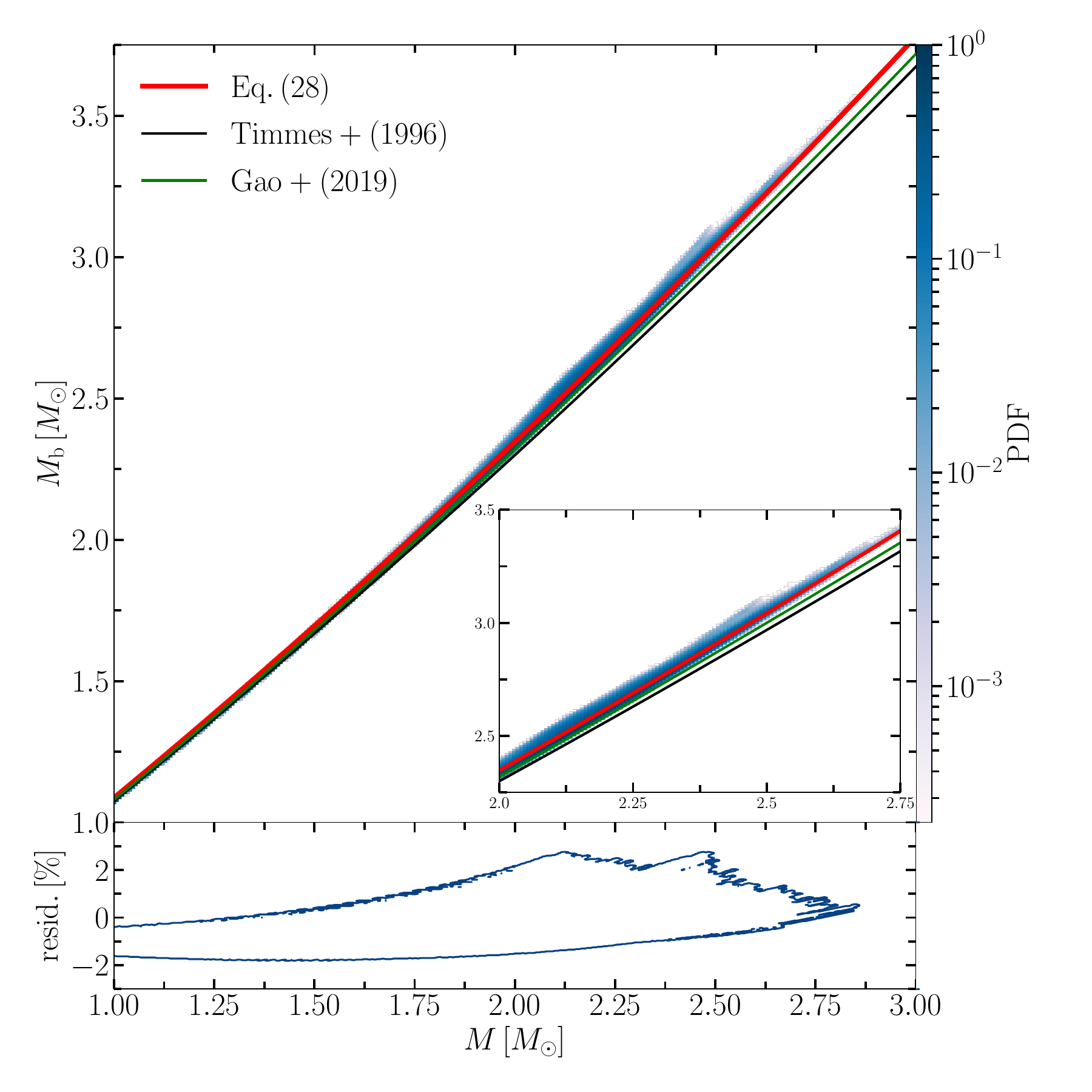}
  \hspace{1.5cm}
  \includegraphics[width=0.4\textwidth]{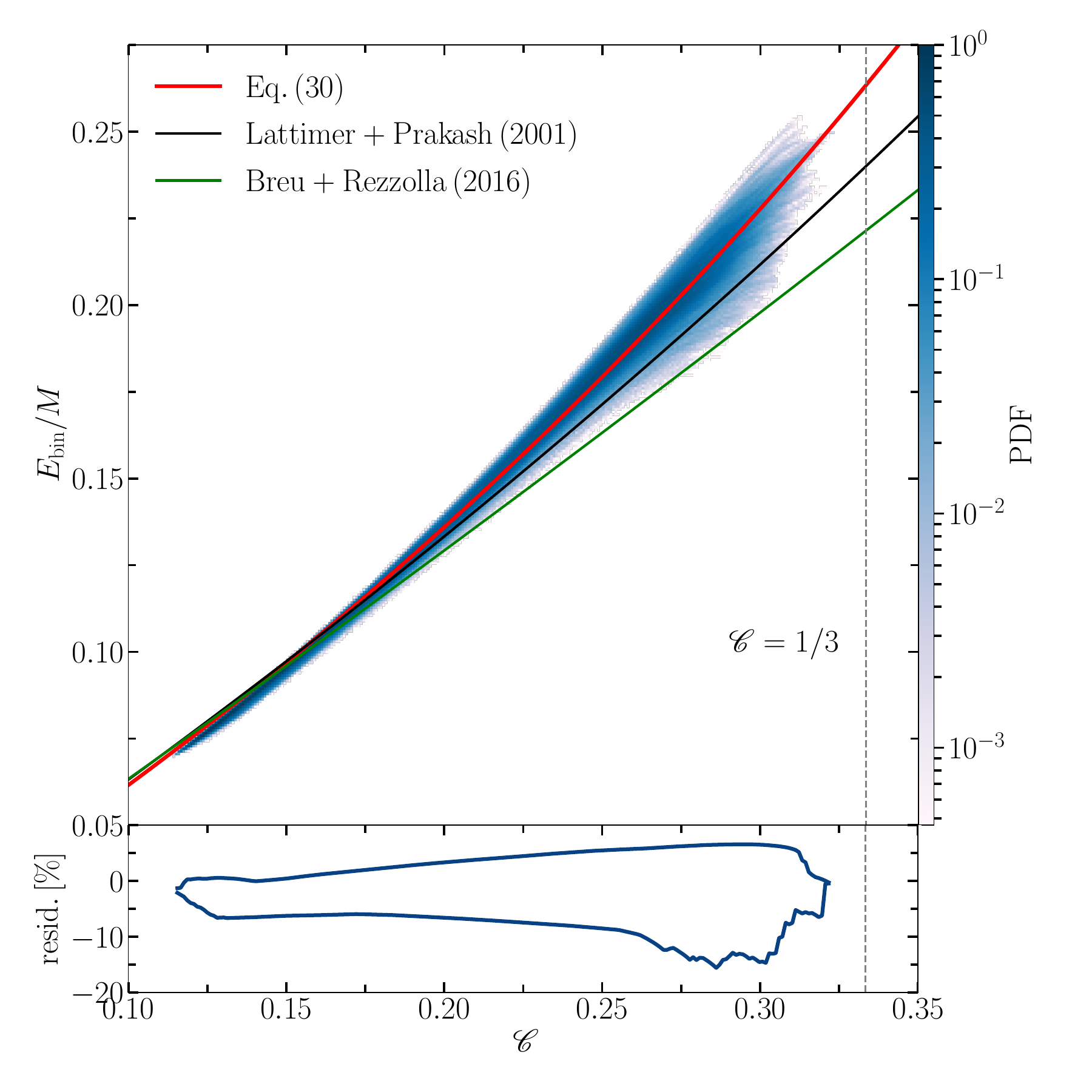}
  \caption{\textit{Left panel:} distribution of the baryonic mass
    $M_\mathrm{b}$ as a function of gravitational mass $M$. Shown
    respectively with red, black and green solid lines are the fits from
    our work, from \citet{Timmes1996}, and from \citet{Gao2020}.
    \textit{Right panel:} distribution of the binding energy $E_{\rm
      bin}/M$ as a function of the stellar compactness
    $\mathscr{C}$. Also in this case, the red line shows our fit to the
    data, while the black and green lines reports the fit
    by~\citet{Lattimer01} and \citet{Breu2016}.}
  \label{fig:fig3}
\end{figure*}

Some models to describe the amount of mass and velocity of ejecta are
parametrized by the gravitational and baryonic mass of the components,
among others~\citep{Coughlin2017}. These models allow us to extract
information from the binary system by comparing observations with
predictions. However, they have degeneracies among these parameters and
others, such as spins of the binary and the EOS of NS matter. As a
result, (quasi-) universal relations can reduce the dimensionality of the
parameter space by trading physical uncertainties against uncertainties
in model assumptions. In the left panel of Fig.~\ref{fig:fig3} we show
the distribution of gravitational mass $M$ vs. baryonic mass
$M_\mathrm{b}$. The solid lines correspond to a quadratic fit of the form
\begin{equation}
\label{eq:Mb_vs_M}
  M_\mathrm{b} = M ( 1 + a_2\, M) \, ,
\end{equation}
as the one presented by \citet{Timmes1996}. In black we show the fit
using a constant $a_2=0.075$ from \citet{Timmes1996}, while in green we
show the fit with $a_2 = 0.080$ from \citet{Gao2020}. Using our set of
EOSs we find that our best fit gives $a_2 = 0.087$. The absolute value of
the residual (using our fit) is $\lesssim 0.08 \, M_\odot$, corresponding
to $\lesssim 3\%$, as shown in the bottom panel. This implies a
quasi-universal relation, since given the gravitational mass of a NS, we
can now predict its baryonic mass with absolute error $\lesssim 0.08 \,
M_\odot$.

\begin{figure}
  \centering
  \includegraphics[width=0.85\columnwidth]{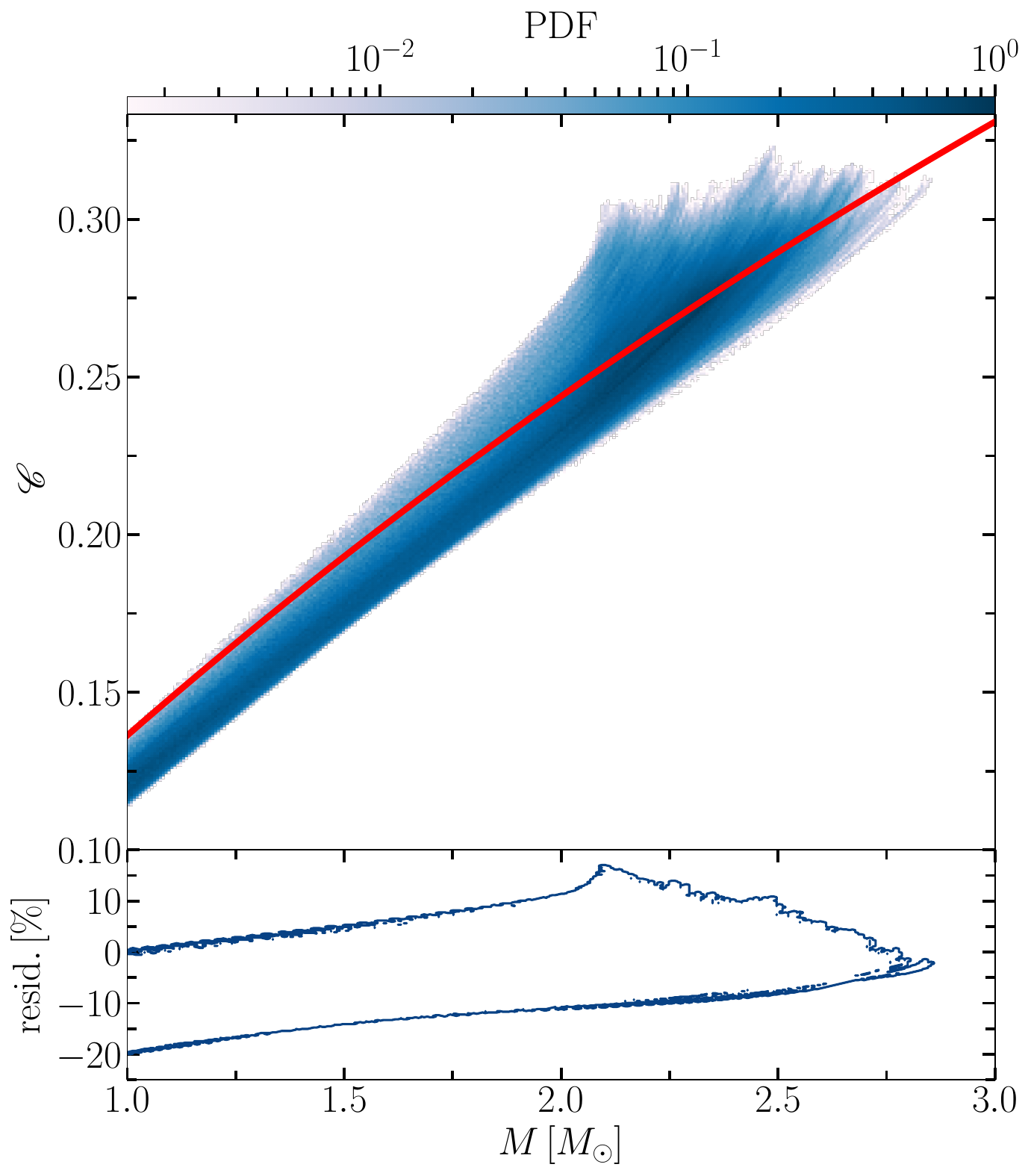}
    \caption{Distribution of the compactness as a function of the
      gravitational mass. Shown with a red solid line is the functional
      behaviour of Eq.~\eqref{eq:comp_vs_M}, which provides a first
      approximation to the compactness, with an uncertainty of
      $15$-$20\%$ for low-mass NSs.}
\label{fig:fig4}
\end{figure}

Exploiting the quasi-universal relation \eqref{eq:Mb_vs_M} it is
straightforward to derive another and \textit{linear} quasi-universal
relation between the binding energy $E_\mathrm{bin}$ and the gravitational
mass
\begin{equation}
\label{eq:Eb_vs_Mb}
  E_\mathrm{bin} := M_\mathrm{b} - M = a_2 M^2 \, ,
\end{equation}
so that a measurement of the gravitational mass of a compact star readily
provides a measure of its binding energy. Furthermore, it has long since
been known that the normalised binding energy $E_\mathrm{bin}/M$ is
linked via another quasi-universal relation to the stellar
compactness~\citep{Lattimer01, Breu2016}
\begin{equation}
  \label{eq:Eb_qu}
  \frac{E_\mathrm{bin}}{M} = \frac{c_1 \mathscr{C}}{1 - c_2
    \mathscr{C}}\,,
\end{equation}
where it is now possible to compute the best-fit values of the
coefficients using the extended set of EOSs and find that $c_1=0.56270
\pm 0.00005$ and $c_2=0.86332 \pm 0.00028$. The result of the fitting is
shown with a red line in the right panel of Fig.~\ref{fig:fig3}, which
reports the distribution of the binding energy $E_{\rm bin}/M$ as a
function of the stellar compactness $\mathscr{C}$. Also in this case, the
red line shows our fit to the data, while the black and green lines
reports the fit by~\citet{Lattimer01} and \citet{Breu2016}\footnote{The
reduced $\chi^2$ obtained by~\citet{Lattimer01} and \citet{Breu2016} for
the old dataset was essentially the same. It is only with the new and
larger dataset that residuals vary.}. Also reported with a vertical black
dotted line is the value $\mathscr{C}=1/3$, which has been recently shown
to represent the upper limit of the compactness for NSs satisfying all
known physical and astrophysical constraints~\citep{Rezzolla2025}.

As a result, once a measurement of the gravitational mass has been made,
the compactness can be estimated as
\begin{equation}
  \label{eq:comp_vs_M}
  \mathscr{C} = \frac{a_2 M}{c_1 + a_2 c_2 M}\,.
\end{equation}
The nonlinear expression~\eqref{eq:comp_vs_M} is shown with a red solid
line in Fig.~\ref{fig:fig4}, which reports the distribution of the
stellar compactness as a function of the gravitational mass. While the
associated uncertainty is not small, it is $\sim 15$-$20\%$ for masses
$\lesssim 2 \, M_{\odot}$ and thus comparable if not smaller than the
observational uncertainty~\citep{Miller2021, Riley2021}.

\begin{figure}
  \centering
  \includegraphics[width=0.95\columnwidth]{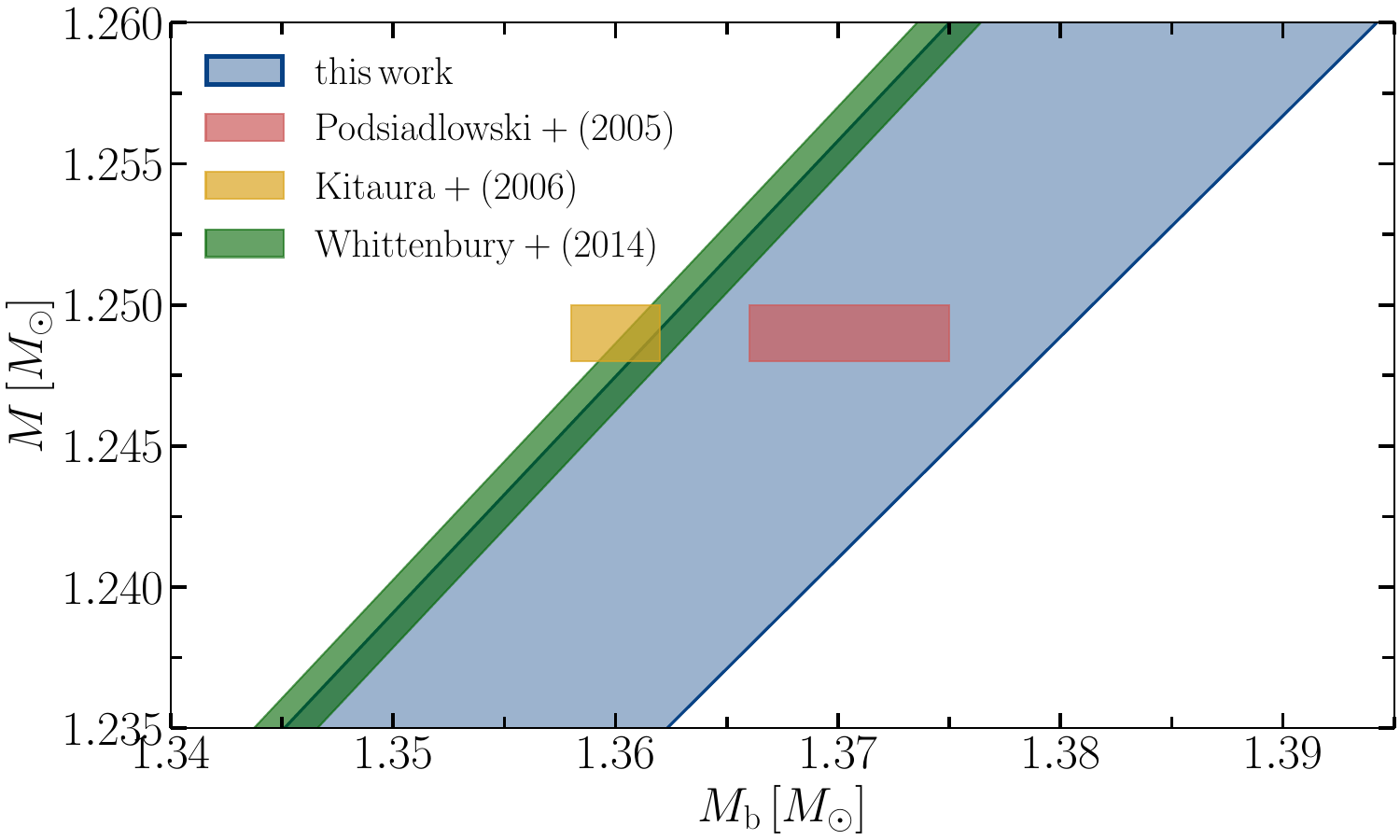}
    \caption{Different estimates of the baryonic mass for pulsar B in the
      binary system J0737-3039. Shown with different shadings are the
      estimates by~\citet{Podsiadlowski2005} (red), \ie $M_{\rm b} =
      1.366 - 1.375\, M_\odot$, by~\citet{Kitaura06} (yellow), \ie
      $M_{\rm b} = 1.360 \pm 0.002\, M_\odot$ and in this work as taken
      from the 100\%-confidence interval of the distribution in the left
      panel of Fig.~\ref{fig:fig3} (blue). The predictions
      by~\citet{Whittenbury2014} are shown as the green stripe.}
\label{fig:fig5}
\end{figure}

The importance of the relation~\eqref{eq:Mb_vs_M} between the baryonic
and the gravitational mass goes beyond being a tight quasi-universal
relation and can actually be employed to obtain an astrophysically
relevant result. More specifically, we recall that measurements on the
gravitational and baryonic mass of the double pulsar J0737-3039 offer a
constraint on the EOS. The gravitational mass of pulsar B in this system
is measured to be $1.2489 \pm 0.0007$ $M_\odot$~\citep{Kramer2006}. Such
low-mass NS may occur for O-Ne-Mg white dwarfs when the core density
reaches a critical value, triggering its collapse. Using this hypothesis
and assuming that the mass loss after the collapse is at negligible (\ie
up to $10^{-2} \,M_\odot$), \cite{Podsiadlowski2005} estimated that the
corresponding baryonic mass of pulsar B lies between $1.366 \, M_\odot$
and $1.375 \, M_\odot$ under the assumption that it was produced in an
electron-capture supernova (red-shaded box in Fig.~\ref{fig:fig5}). On
the other hand, \cite{Kitaura06} used 1D simulations of type II supernova
explosions and estimated a smaller baryonic mass of the resulting NS of
$M_\mathrm{b} = 1.360 \pm 0.002$ $M_\odot$ (yellow-shaded box in
Fig.~\ref{fig:fig5}), due to loss in baryonic mass during the
explosion. The 100\%-confidence interval for the baryonic mass of pulsar
B inferred from our set of EOSs is given by the blue-shaded area between
the two horizontal dashed lines indicating the uncertainty in the
measurement of the gravitational mass~\citep{Kramer2006}. The two slanted
blue solid lines limiting the blue-shaded area are obtained after
inverting Eq.~\eqref{eq:Mb_vs_M} and correcting for the spread in the
correlation between the two masses. The corresponding analytic
expressions are therefore\footnote{These analytic expressions are
accurate in the range $M_\mathrm{b}/M_{\odot} \in [1.3, 2.3]$.} $M_{l,u}
= ({\sqrt{1 + 4 (b_{1})_{l,u} M_\mathrm{b}}} - 1) / {2 (b_{1})_{l,u}} +
(b_{2})_{l,u}$, where the coefficients are given by $(b_{1})_l = 0.1139,
(b_{2})_l = 0.0364$ and $(b_{1})_u = 0.0791, (b_{2})_u = 0.0088$, where
''$l$'' and ''$u$'' refer to the lower and upper confidence limits,
respectively.  Finally, the interval obtained by
Ref.~\cite{Whittenbury2014} is shown as the green-shaded stripe, for
comparison. Our results show consistency with the whole baryonic mass
range estimated by \cite{Podsiadlowski2005}, while the range $M_{\rm b} =
1.3596 \,- 1.3620 \, M_\odot$ is compatible with the estimate
of~\cite{Kitaura06}.

\begin{figure*}
\centering
\includegraphics[width=0.49\textwidth]{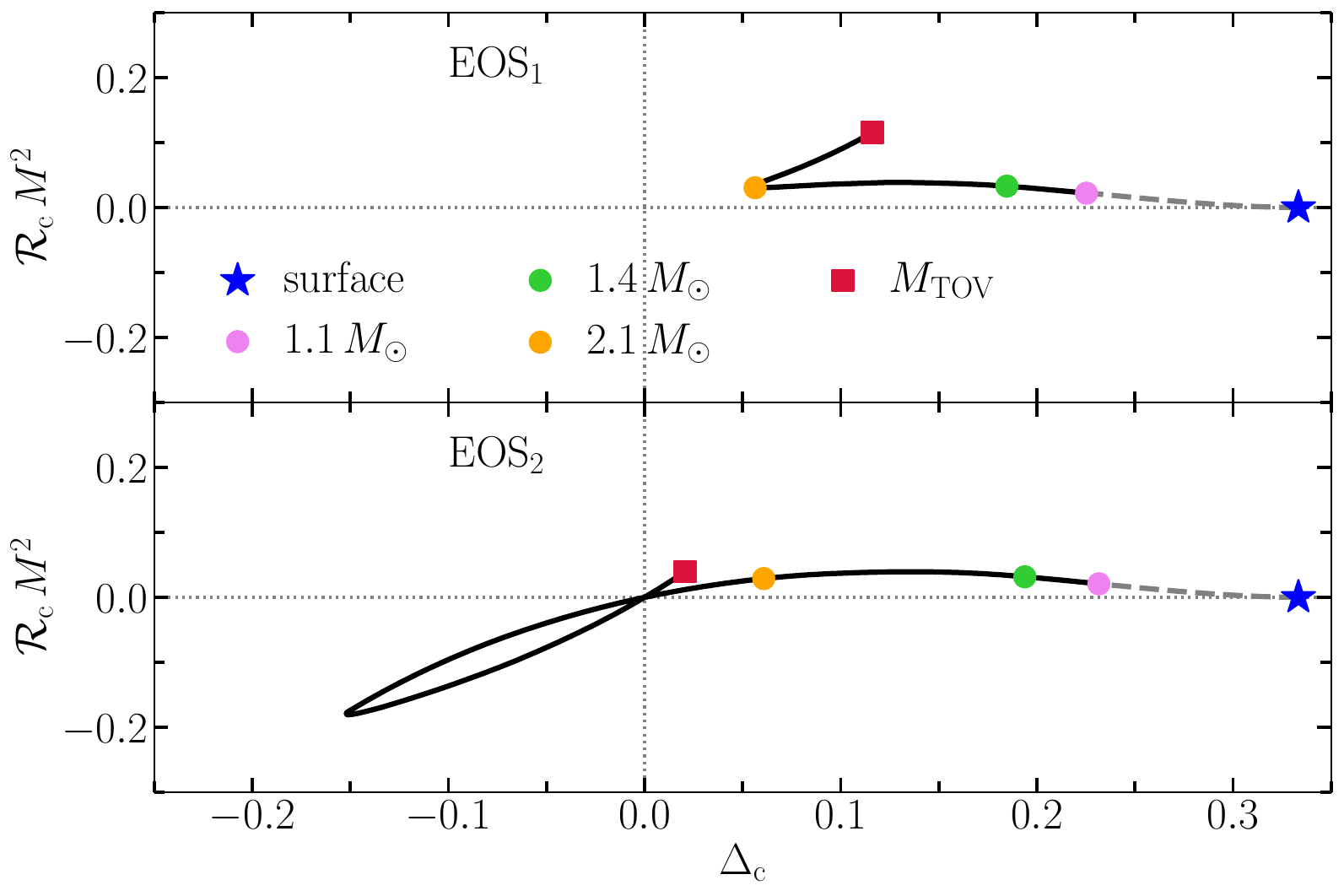}
\hspace{0.125cm}
\includegraphics[width=0.49\textwidth]{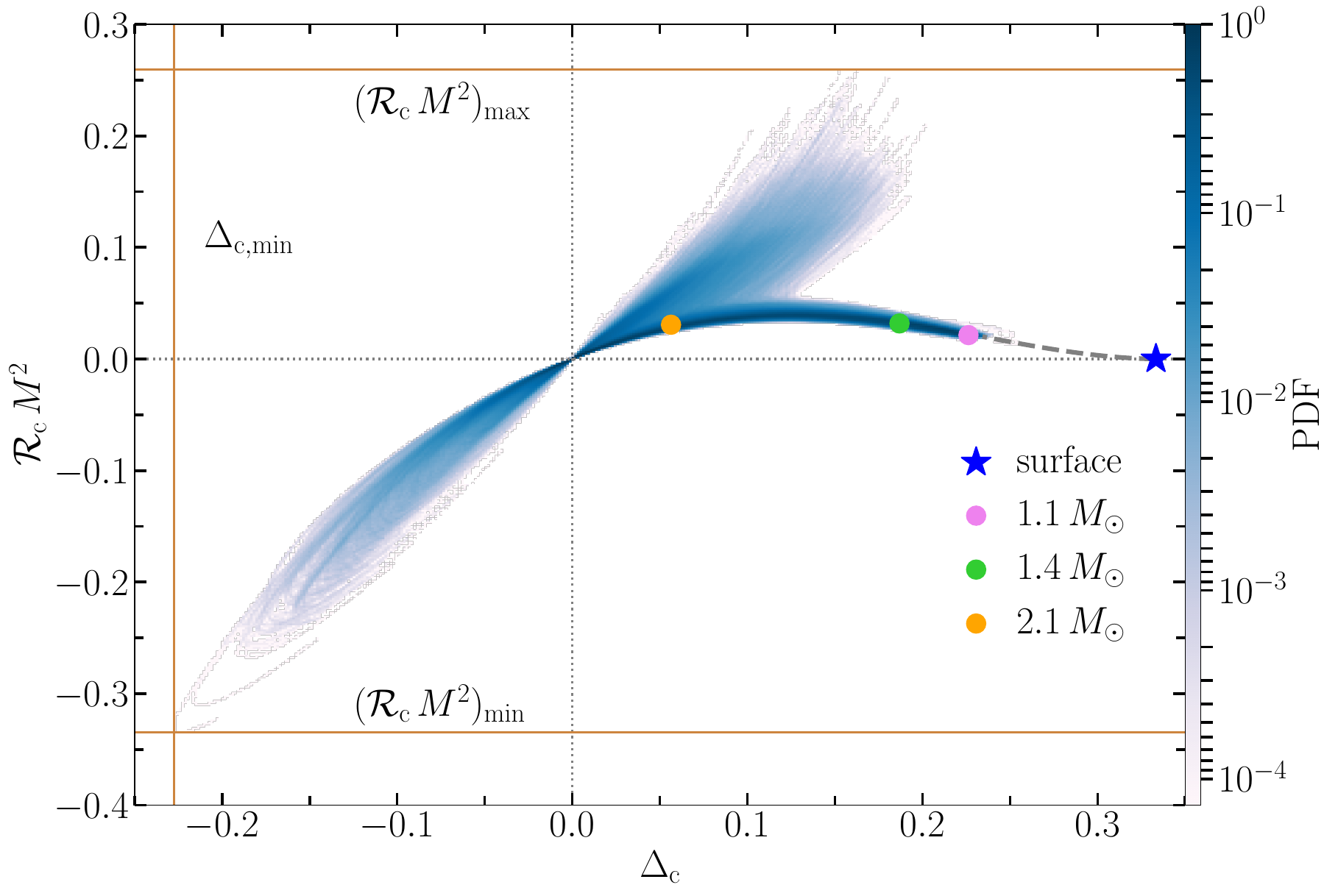}
\caption{\textit{Left panel:} relation between the trace anomaly and the
  Ricci scalar for two representative EOSs: ${\rm EOS}_1$ (top part) and
  ${\rm EOS}_2$ (bottom part). For each curve, the relation will be
  represented by a curve that starts at the surface of the star where
  $\Delta_\mathrm{s} = 1/3 , \, \mathcal{R}_\mathrm{s} = 0$ (blue
  asterisk) and ends at the centre of the star where $\Delta_\mathrm{c} ,
  \mathcal{R}_\mathrm{c}$ (coloured filled circles); the grey dashed line
  is used to mark stars with $M\leq 1.1\,M_{\odot}$, that are
  astrophysically unlikely, while the red squares mark the largest
  possible mass. Depending on the EOS, the black curve may not reach
  $\Delta_\mathrm{c}=0$ (as for the case of ${\rm EOS}_1$), or reach it
  and perform a loop in the negative $(\mathcal{R}_c, \Delta_\mathrm{c})$
  region (as for the case of ${\rm EOS}_2$). \textit{Right panel:} the
  same as in the left but shown for all the EOSs considered in terms of a
  normalised PDF. Also shown with vertical and horizontal brown lines
  are, respectively: the minimum value of central conformal anomaly
  $\Delta_\mathrm{c, min}$, and the minimum (maximum) value of normalised
  central value of Ricci scalar $(\mathcal{R} _\mathrm{c} \,
  M^2)_\mathrm{min}$ ($(\mathcal{R} _\mathrm{c} \, M^2)_\mathrm{max}$).}
\label{fig:fig6}
\end{figure*}

\subsection{Ricci Curvature and Trace Anomaly}

The radial profiles $e=e(r)$ and $p=p(r)$ obtained by solving the TOV
equations allow us to compute the relationship between the conformal
anomaly $\Delta(r)$ and the Ricci scalar $\mathcal{R} (r)$, via
Eq.~\eqref{eq:Ricci}. However, $\mathcal{R} (r)$ and $\Delta (r)$ are
simply proportional, hence, the radial dependence of the latter would be
as that discussed for the formed in Fig.~\ref{fig:fig1} scaled with the
energy density. As a result, it is more interesting to study here the
correlation between the two scalar functions. What needs to be borne in
mind is that, once an EOS is fixed, for a NS with mass $M$ and radius
$R$, the relation between the trace anomaly and the Ricci scalar will be
given by a curve that starts at $\left( \Delta _\mathrm{s} = 1/3 , \,
\mathcal{R}_\mathrm{s} = 0 \right)$ -- the surface of the star (blue
asterisk in the lefts panel of Fig.~\ref{fig:fig6}) -- and ends at some
point $\left(\Delta_\mathrm{c} , \mathcal{R}_\mathrm{c} \right)$ -- the
centre of the star (coloured filled circles in the left panel of
Fig.~\ref{fig:fig6}); the grey dashed line is used to mark stars with
$M\leq 1.1\,M_{\odot}$, that are astrophysically unlikely. Clearly, the
endpoint of such a curve will depend on the value of $M$ and will
eventually stop for the largest mass possible for each EOS (red squares
in the left panel of Fig.~\ref{fig:fig6}). Interestingly, this behaviour
not only is non-monotonic, but it can also be double-valued both in the
trace anomaly and the Ricci scalar. Furthermore, since we have discussed
that $\Delta$ and $\mathcal{R}$ generally move towards negative values
for the most compact and most massive stars, the behaviour of the
normalised Ricci scalar relative to the trace anomaly can be of the type
shown in the lower-left panel of Fig.~\ref{fig:fig6}, where the curve
passes twice through the point $\left( \Delta = 0, \, \mathcal{R} = 0
\right)$ for a given ``$\mathrm{EOS}_2$''. This EOS leads to very massive
and compact stars (as can be seen in the left panel of
Fig.~\ref{fig:fig2}) and hence encounters negative values of the trace
anomaly and of the Ricci scalar somewhere inside the corresponding
stellar models. This behaviour should be contrasted with that exhibited
by a different EOS, \ie ``$\mathrm{EOS}_1$'' in the upper-left panel of
Fig.~\ref{fig:fig6}, which does not produce negative values.

The right panel of Fig.~\ref{fig:fig6}, shows with a colormap the PDF of
the normalised Ricci scalar as a function of the trace anomaly, at the
centre of the stars, for all the EOSs considered in our sample. Clearly,
the point $(\Delta_\mathrm{s} = 1/3, \, \mathcal{R}_\mathrm{s} = 0)$ is
the starting point for all the EOSs and all curves have to pass through
it. However, depending on the EOS, the $\mathcal{R}-\Delta$ curve may or
may not pass through the point $(\Delta = 0, \, \mathcal{R} = 0)$, and
may end at either $\mathcal{R}_\mathrm{c} < 0$, $\mathcal{R}_\mathrm{c} =
0$, or $\mathcal{R}_\mathrm{c} > 0$. Marked with solid brown lines are
the minimum [maximum] normalised Ricci scalar $(\mathcal{R} _\mathrm{c}
\, M^2)_\mathrm{min}$ [$(\mathcal{R} _\mathrm{c} \, M^2)_\mathrm{max}$]
and the minimum trace-anomaly $\Delta _\mathrm{c,min}$ found with our set
of EOSs.

The global minimum value of the Ricci scalar achieved inside any NS is
found to be $\mathcal{R}_\mathrm{min} = -2.510 \times 10^{-12}
\mathrm{cm}^{-2}$; while the global maximum value is $\mathcal{R}
_\mathrm{max} = 2.383 \times 10^{-12} \mathrm{cm}^{-2}$. The global
minimum of the trace anomaly turns out to be $\Delta_\mathrm{min} =
-0.227$. These three values can be taken as bounds for both the Ricci
scalar and the trace anomaly present in the most compact objects
composed of regular matter in the Universe.

We conclude this Section by commenting on the result that the Ricci
scalar can achieve negative values in the interior of massive stars.
First, we note that a negative value of the Ricci scalar does not reflect
a violation of any of the energy
conditions~\citep{Rezzolla_book:2013}. Indeed, since the energy density
and pressure are always positive in all of our EOSs, the weak (\ie $e + p
\geq 0$, $e \geq 0$) and strong (\ie $e + p \geq 0$, $e + 3p \geq 0$)
energy conditions are satisfied for all of them. In addition, the
dominant energy condition (\ie $e \geq |p|$), can be recast into a
condition on the trace anomaly, \ie $\Delta > -2/3$, which is clearly
satisfied by all of the EOSs (see right panel of
Fig.~\ref{fig:fig6}). Second, negative values of the Ricci scalar may
appear somewhat surprising because intuition suggests that curvature
should always be positive in the star and actually decrease almost
monotonically\footnote{The strict monotonicity is broken at the stellar
surface because of the contributions coming from $\mathcal{I}_1$; see
also~\cite{Eksi2014}.} from the centre. This intuition is indeed correct
but does not apply to the Ricci scalar but, rather, to the Kretschmann
scalar. Third, for the static fluid configurations considered here,
\hbox{$\mathcal{R}_{\mu\nu} u^{\mu} u^{\nu} > 0$} everywhere, so that --
using the Raychaudhuri equation -- timelike geodesics would still be
focused even in regions where $\mathcal{R} <
0$~\cite{Rezzolla_book:2013}.

\begin{figure}
  \centering
  \includegraphics[width=0.95\columnwidth]{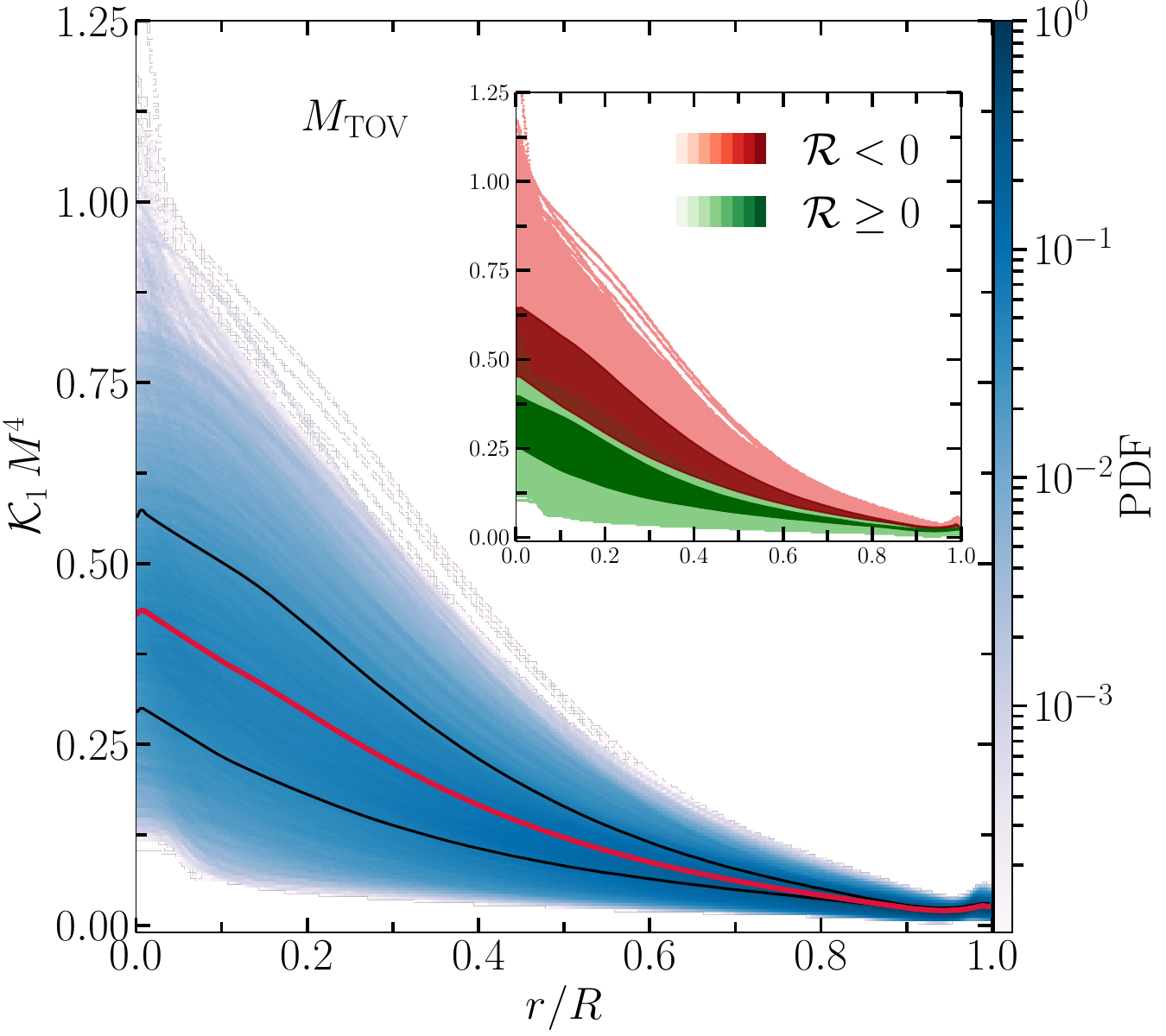}
  \caption{Normalised PDFs of the radial profiles of the normalised
    Kretschmann scalar $\mathcal{K}_1 \, M_{}^4$ for maximum-mass
    stars. Also in this case, the red lines mark the median values of the
    distribution, while black lines correspond to $1$-$\sigma$ confidence
    limits. Shown instead in the inset are the same distributions as in
    the main panel but with red (green) colormaps to mark the regions
    where the NSs have $\mathcal{R}<0$ ($\mathcal{R}\geq0$) somewhere in
    their interior. Darker/lighter shadings refer to $1$-$\sigma$ and
    100\% confidence levels, respectively.}
  \label{fig:fig7}
\end{figure}

\begin{figure*}
  \centering
  \includegraphics[width=0.32\textwidth]{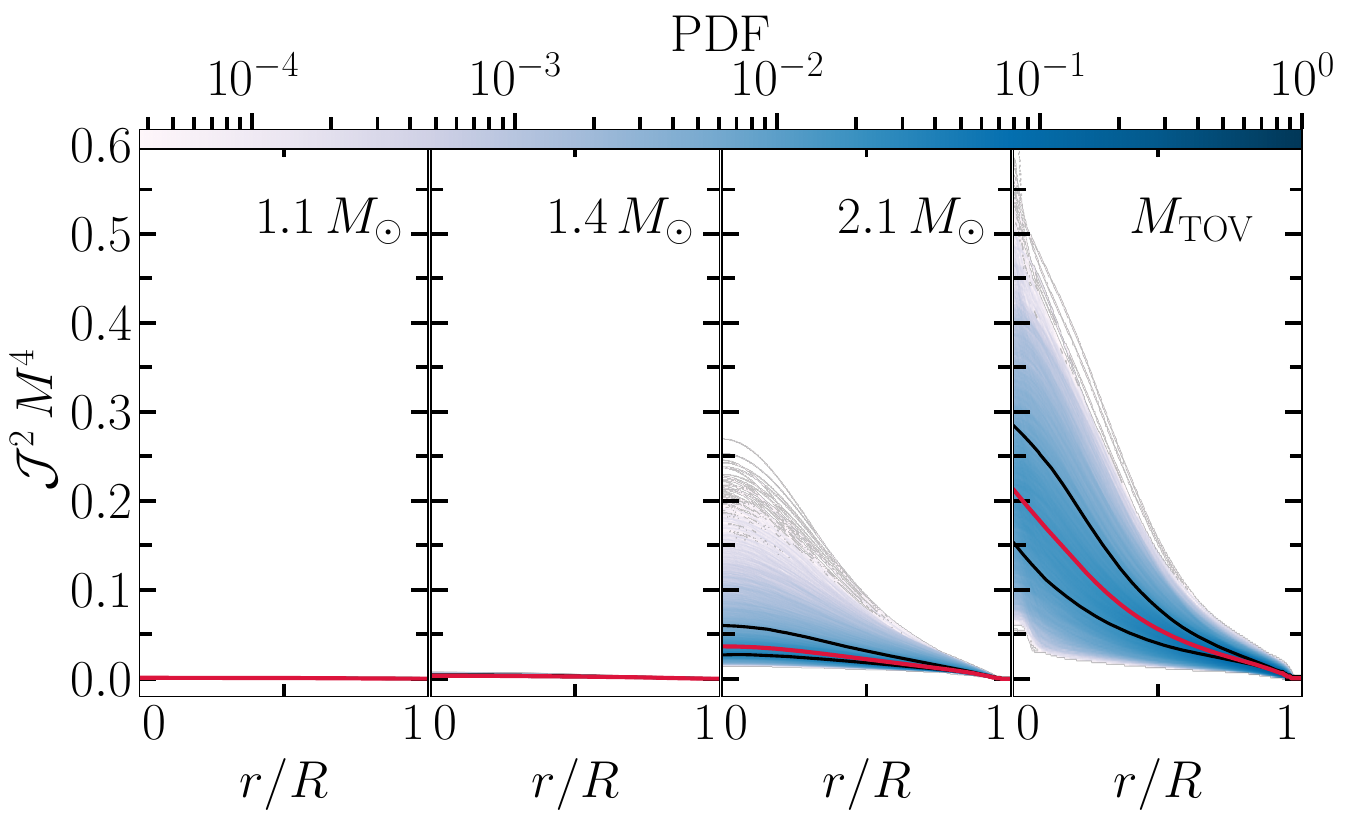}
  \hspace{0.1cm}
  \includegraphics[width=0.32\textwidth]{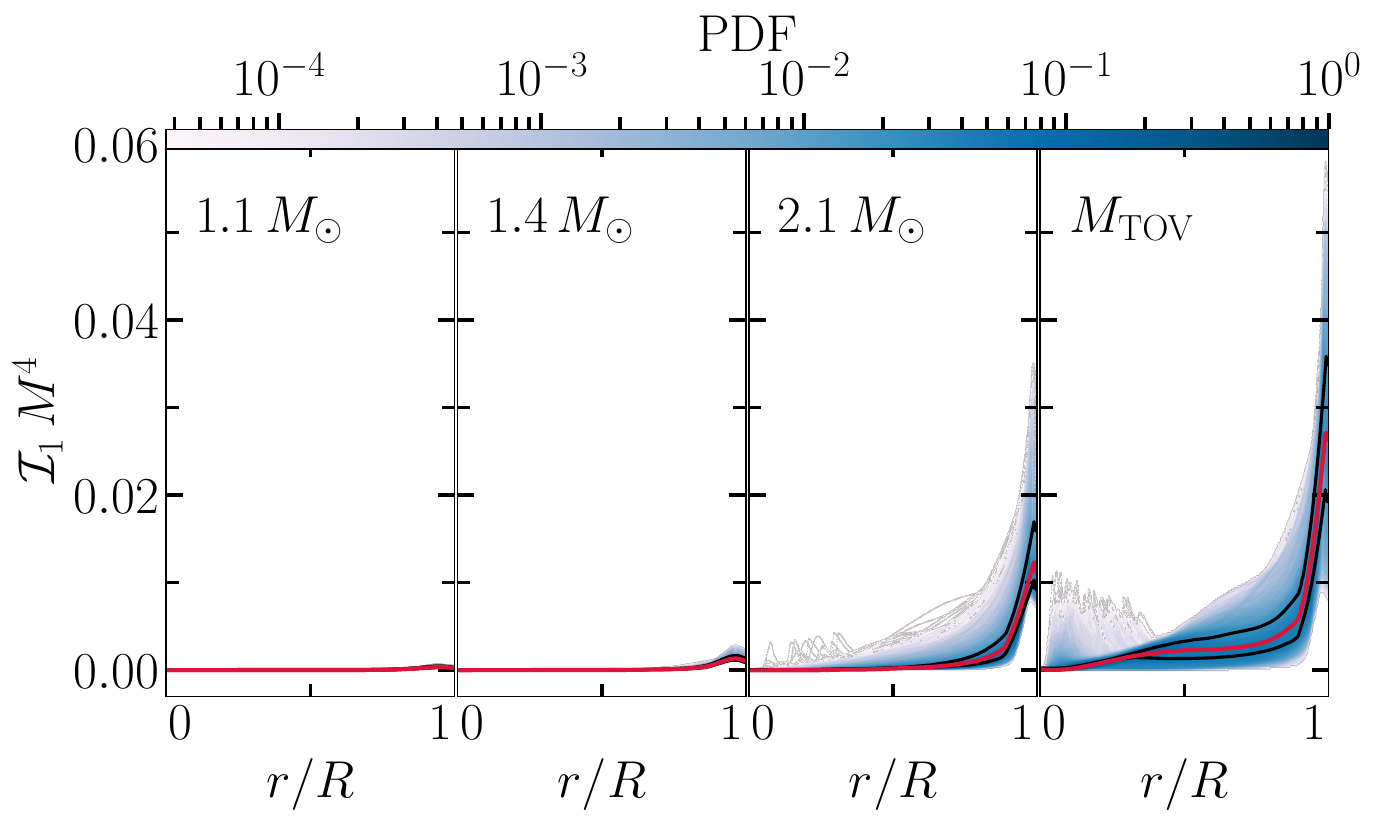}
  \hspace{0.1cm}
  \includegraphics[width=0.32\textwidth]{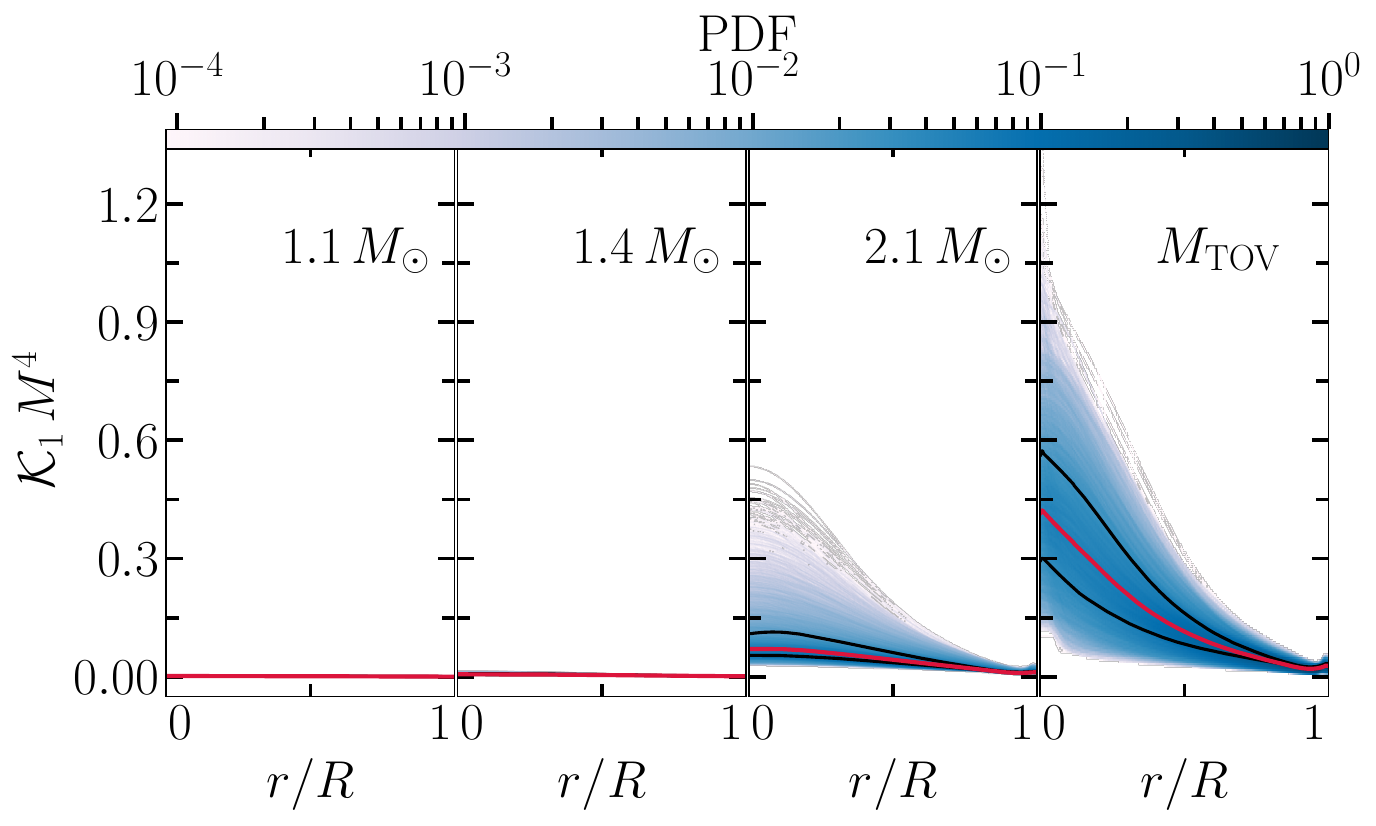}\\
  \includegraphics[width=0.32\textwidth]{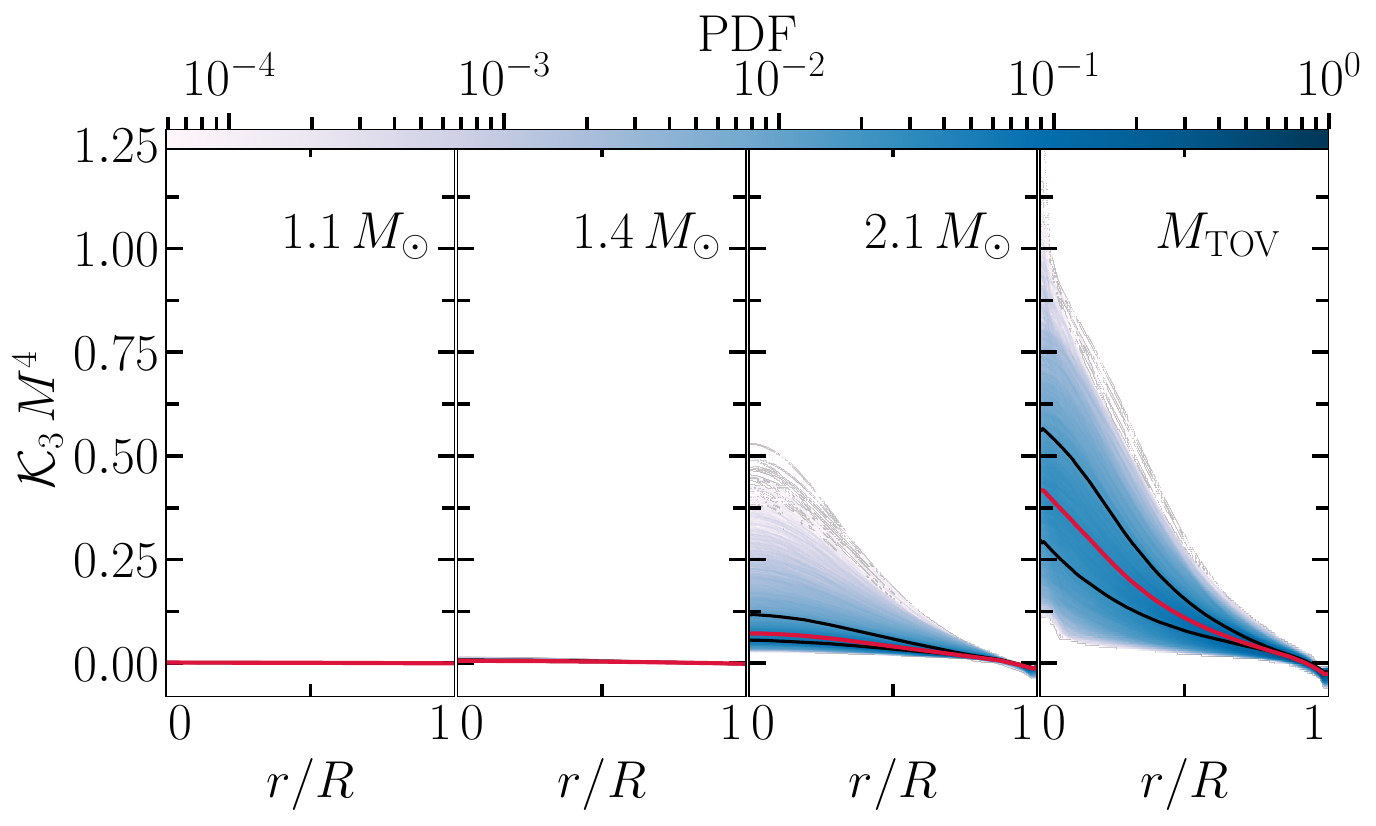}
  \hspace{0.1cm}
  \includegraphics[width=0.32\textwidth]{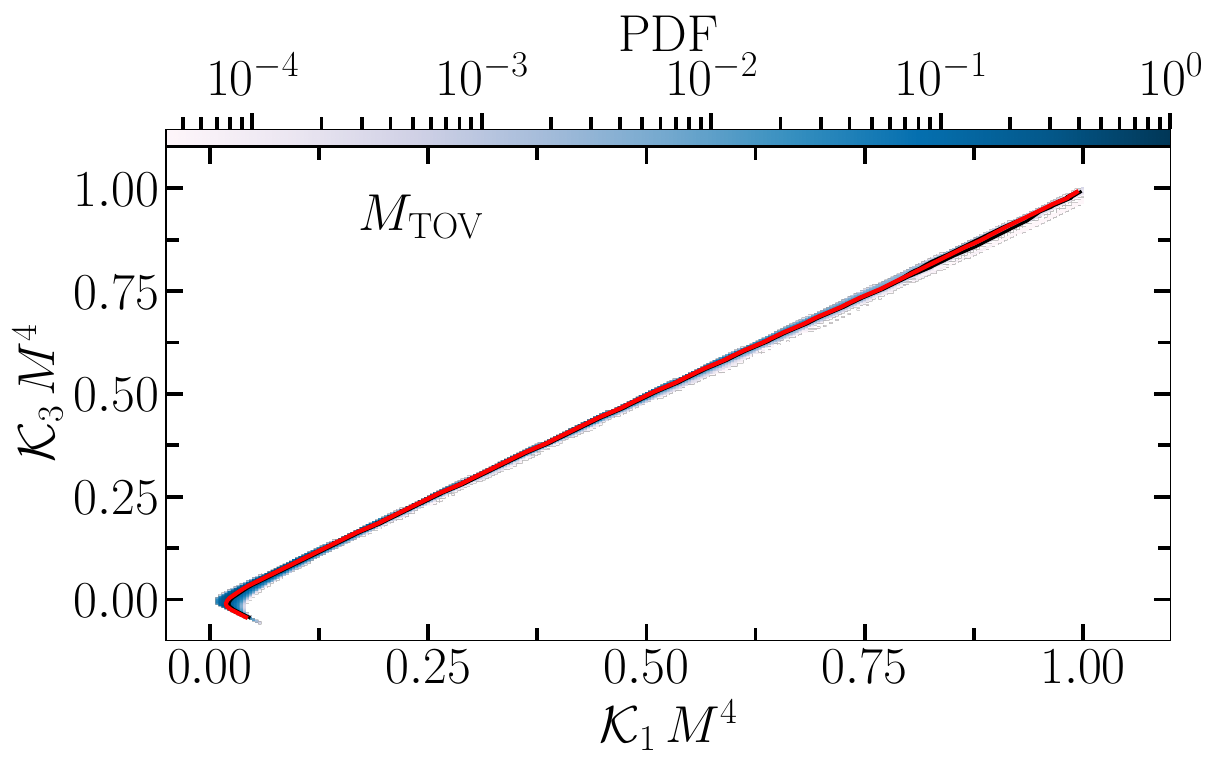}
  \hspace{0.1cm}
  \includegraphics[width=0.32\textwidth]{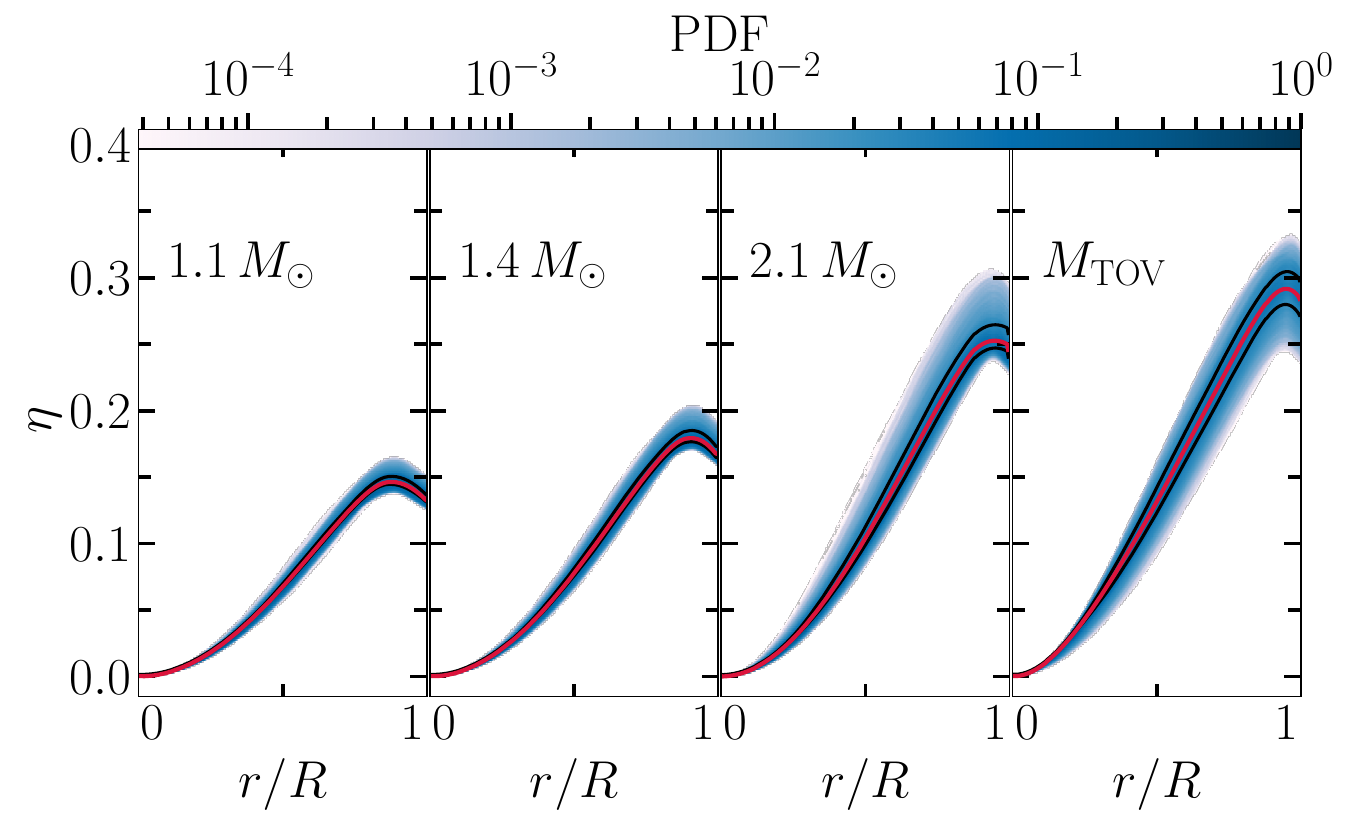}
  \caption{Normalised PDFs of the radial profiles of various normalised
    curvature scalars. From top left, we report $\mathcal{J}^2$,
    $\mathcal{I}_1$, $\mathcal{K}_1$, and $\mathcal{K}_3$ properly
    rescaled by a factor $M^4$ to make them dimensionless, and for the
    same reference masses of $M=1.1$, $1.4$, $2.1 \, M_\odot$, and
    $M_{_{\mathrm{TOV}}}$. The fifth panel from the top left highlights
    the close correlations between $\mathcal{K}_3$ and $\mathcal{K}_1$,
    while the last panel shows the radial profile of the compactness
    function, $\eta(r) := m(r)/r$. Also here, the red lines mark the
    median values of the distribution, while black lines correspond to
    $1$-$\sigma$ confidence limits.}
  \label{fig:fig8}
  \end{figure*}

Figure~\ref{fig:fig7} provides an analogue to the rightmost panel of
Fig.~\ref{fig:fig1} but for the dimensionless Kretschmann scalar, \ie
$\mathcal{K}_1\,M^4$, and for stellar models with $M=M_{_{\rm
    TOV}}$. Note that in this case $\mathcal{K}_1\,M^4$ is always
positive and decreasing outwards, highlighting that the Ricci scalar is
simply not a very representative indicator of the curvature in a star as
it is not in a Schwarzschild (Kerr) black hole, since it is zero
everywhere apart from the singularity where it is ill-defined. The
Kretschmann scalar, on the other hand, not only respects the expected
behaviour (which is similar to that in a black hole) but it is also
larger for more compact and massive stars. This is illustrated in the
inset of Fig.~\ref{fig:fig7} which reports with a green (red) colourmap
the Kretschmann scalar for stellar models where $\mathcal{R} > 0$
everywhere ($\mathcal{R} \leq 0$ somewhere) inside the star, and where
the dark and light shadings cover the 1-$\sigma$ and 100\%-confidence
intervals, respectively. Finally, we note that the value of
$\mathcal{K}_1 \,M^4_{_{\rm TOV}}$ at the centre of the star is directly
correlated with the corresponding value of $\mathcal{R} \,M^2_{_{\rm
    TOV}}$, i.e., is larger for those stellar models which have the
smallest and negative values of $\mathcal{R} \,M^2_{_{\rm TOV}}$. This
behaviour is not obvious to expect from Eq.~\eqref{eq:K_1}, which relates
$\mathcal{K}_1$ to $\mathcal{R}$ and $\mathcal{J}^2$ ($\mathcal{I}_1 = 0$
at the centre of the star; see Fig.~\ref{fig:fig8}). Yet, as the stellar
centre is approached, $2\mathcal{J}^2$ grows more rapidly than
$\mathcal{R}^2/3$, hence ensuring that $\mathcal{K}_1$ not only is
positive, but it is also larger for more massive stars.

\subsection{Other Curvature invariants}
\label{sec:other_inv}

Figure~\ref{fig:fig8} provides information that is complementary to that
in Fig.~\ref{fig:fig1}, namely, it reports the radial profiles of the
other curvature scalars described in Section~\ref{sec:intro}. Starting
from the top left, we report the PDFs for $\mathcal{J}^2$,
$\mathcal{I}_1$, $\mathcal{K}_1$, and $\mathcal{K}_3$ properly rescaled
by a factor $M^4$ to make them dimensionless, and for the same reference
masses of $M=1.1$, $1.4$, $2.1 \, M_\odot$, and $M_{_{\mathrm{TOV}}}$.

As expected from Eq.~\eqref{eq:J}, the full contraction of the Ricci
tensor, $\mathcal{J}^2$, decreases from the centre to the stellar surface
and because $\mathcal{J}^2$ is proportional to the squared energy density
$e^2$. In this respect, this scalar is similar to the Ricci scalar
$\mathcal{R}$, since it is related to the part of the gravitational field
that is tied directly and locally to the matter sources of the field. On
the other hand, the full contraction of the Weyl tensor, $\mathcal{I}_1$,
depends on the squared difference between the mean energy density and the
energy density, $(\bar{e}-e)^2$. While close to the centre of the NS,
$\bar{e} \sim e$ and $\mathcal{I}_1 \sim 0$, with increasing radial
coordinate, the inequality $\bar{e} > e$ becomes steeper. As a result, at
lower densities and close to the surface, the relevant quantity for
$\mathcal{I}_1$ is $\bar{e} = {3m}/({{4} \pi r^3})$, so that
$\mathcal{I}_1$ increases until it reaches a maximum near the surface and
then decreases. This decreasing behaviour is due to the fact that the
integrated mass increases at a slower rate than $\propto r^3$ as it
approaches the surface. Because outside the NS it will decrease as
$\propto {M^2}/{r^6}$ (not shown in Fig.~\ref{fig:fig8}), we can
interpret the scalar $\mathcal{I}_1$ as the curvature produced by the
gravitational field of the mass distribution $m(r)$ or, alternatively, as
the deviation of the local curvature at a given radial position and from
the average one up to that position.

On the other hand, when considering the radial profiles of the
Kretschmann scalar, $\mathcal{K}_1$, the PDFs show the behaviour dictated
by Eq.~\eqref{eq:K_1}, so that at the core of the NSs, this scalar is
mainly determined by $2\mathcal{J}^2$, and close to the surface it
approaches $\mathcal{I}_1$; in the vacuum outside the NSs (not shown in
Fig.~\ref{fig:fig6}), $\mathcal{K}_1 =\mathcal{I}_1$. Thus,
$\mathcal{K}_1$ can be thought to be associated with both the local
energy distribution $e(r)$ and with the mass distribution $m(r)$, much as
in a Newtonian gravitational potential.

Finally, when exploring the radial behaviour of the Euler scalar
$\mathcal{K}_3$, it is remarkable to note the similarities, both
qualitative and quantitative, with the Kretschmann scalar
$\mathcal{K}_1$. This is in the middle panel of the second row in
Fig.~\ref{fig:fig8}, which reports the PDF of $\mathcal{K}_3$ as a
function of $\mathcal{K}_1$. It is then apparent that using
Eqs.~\eqref{eq:K_1} and~\eqref{eq:K_3} and with the exception of the
stellar surfaces -- where $\mathcal{K}_1 \simeq 0$ -- we infer
\begin{align}
& \mathcal{K}_3 \sim \mathcal{K}_1 \sim 2\mathcal{J}^2 \,,\\
& 2\mathcal{J}^2 \gg \mathcal{I}_1 - \mathcal{R}/3\,,\\
& 2\mathcal{J}^2 \gg - \mathcal{I}_1 - 2\mathcal{R}/3\,.
\end{align}
Near the stellar surfaces, on the other hand, the term $-\mathcal{I}_1$
dominates, so that at the lowest densities $\mathcal{K}_3 \approx
-\mathcal{K}_1 \lesssim 0$. The fact that the Euler scalar can reach
negative values near the stellar surface makes it less useful than the
Kretschmann scalar, even though it is very similar to the former for most
of the stellar structure. Finally, reported in the bottom-right panel of
Fig.~\ref{fig:fig8}, is the radial profile of the compactness function,
$\eta(r) := m(r)/r$, which exhibits an increasing behaviour in the
stellar core, until a local maximum is produced by a growth in the mass
function that is slower than $\propto r$, thus producing a decrease of
$\eta$ as it approaches the stellar surface; this behaviour was already
remarked by~\citet{Eksi2014}.

\section{Conclusion}
\label{sec:conclusion}

Motivated by the lack of information about how the curvature varies
inside NSs and how it depends on different EOSs, we have performed the
first systematic analysis of curvature invariants in these compact
objects. To this end, we have employed the approach presented
by~\citet{Altiparmak:2022} and constructed a set of $10^4$ EOSs
parameterized in term of the sound speed and that satisfy the constraints
from nuclear theory and perturbative QCD, as well as measurements of NS
masses, radii, and gravitational waves from binary NS mergers.

In this way, and considering the principal curvature invariants of the
Riemann tensor (\eg the Kretschmann, Chern-Pontryagin, Euler invariants),
it was possible to show that NSs can have a negative value of the Ricci
scalar $\mathcal{R}$ somewhere in their interior. Indeed, this property
is so common that about $\sim 50\%$ of our EOSs produce one or more stars
with negative Ricci curvature inside the star. The negative curvature is
found mostly but not exclusively at the highest densities and pressures,
and predominantly for stiff EOSs and for the most compact and most
massive stars. Furthermore, we can calculate a statistically improved
estimate of the ``zero-curvature'' compactness relation, finding that the
value of the compactness $\mathscr{C}:=M/R$ at which the stellar model
has zero Ricci curvature at the center, finding it differs by only $\sim
3\%$ from previous estimates obtained with a much smaller set of
EOSs~\citep{Podkowka2018}.

A negative Ricci scalar may appear counter-intuitive as one may be
induced to think that curvature should always be positive in the star and
actually decrease almost monotonically from the centre. In practice, this
result simply highlights that the Ricci scalar is not a very
representative indicator of the curvature in a star, just as it is not in
a Schwarzschild or a Kerr spacetime. Instead, a much better
indicator of the curvature is the Kretschmann scalar $\mathcal{K}_1$ (and
its related Euler scalar $\mathcal{K}_3$) as it possesses all the
expected properties of positivity and monotonicity.

Our analysis of the general gravitational properties of NSs has also
considered the well-known quasi-universal relation between the stellar
gravitational mass $M$ and the baryonic mass $M_\mathrm{b}$, and we have
derived a new and improved analytic fit of the data that reproduces it
very closely, with a maximum variance of only $\sim 3\%$ across the whole
set of EOSs. As a practical application, we use the new quasi-universal
expression to obtain a new estimate of the baryonic mass of the double
pulsar J0737-3039.

Finally, using the relation between the Ricci scalar and the trace
anomaly $\Delta$, we determine under which conditions $\Delta$ vanishes
or becomes negative in NSs, and also set general upper and lower bounds
on the values of the Ricci scalar at the stellar centre and, more
importantly, the largest violation (\ie the largest negative value) of
the trace anomaly in the interior of a NS. While our results have been
confined to nonrotating stars, very little is presently known about the
curvature behaviour in compact rotating stars, especially when their spin
is near the maximal one. Similarly, while it is difficult to
  predict whether the behaviour found for these scalars will be different
  in other theories of gravity, it certainly represents an interesting
  question. We plan to investigate these aspects in future work.

\section*{Acknowledgements}

We thank Dany Page, Sergio Mendoza, Yavuz Eksi for insightful
discussions, and Mabel Osorio-Archila for useful discussions and careful
reading of the original manuscript. Partial funding comes from the ERC
Advanced Grant ``JETSET: Launching, propagation and emission of
relativistic jets from binary mergers and across mass scales'' (Grant
No. 884631). I.~G.  acknowledges that this work was made possible with
the support of a scholarship from the German Academic Exchange Service
(DAAD), through the Research Grants - Doctoral Programmes in Germany,
personal reference number 91940015. C.~E. acknowledges support by the
Deutsche Forschungsgemeinschaft (DFG, German Research Foundation) through
the CRC-TR 211 ``Strong-interaction matter under extreme conditions''--
project number 315477589 -- TRR 211. L.~R. acknowledges the Walter
Greiner Gesellschaft zur F\"orderung der physikalischen
Grundlagenforschung e.V. through the Carl W. Fueck Laureatus Chair. The
calculations were performed in part on Universidad Nacional Aut\'onoma de
M\'exico's server Geminga and on the ITP Supercomputing Cluster Calea.

\section*{Data Availability}
Data is available upon reasonable request from the corresponding author.

\bibliographystyle{apsrev4-1} 
\bibliography{aeireferences.bib}

\end{document}